\documentclass[preprint2]{aastex63}
\usepackage{graphicx}
\usepackage{psfig}
\usepackage{natbib}
\usepackage{subfigure} 

\newcommand{\gsim}{\mbox{\hspace{.2em}\raisebox{.5ex}{$>$}\hspace{-.8em}\raisebox{-.5ex}{$\sim$}\hspace{.2em}}}
\newcommand{\lsim}{\mbox{\hspace{.2em}\raisebox{.5ex}{$<$}\hspace{-.8em}\raisebox{-.5ex}{$\sim$}\hspace{.2em}}}

\newcommand{\E}[1]{\times 10^{#1}}
\newcommand{\twCO}{$^{12}$CO}  \newcommand{\thCO}{$^{13}$CO}

\newcommand{\HII}{\mbox{H\,\textsc{ii}}}
\newcommand{\HI}{\mbox{H\,\textsc{i}}}
\newcommand{\CII}{\mbox{C\,\textsc{ii}}}

\newcommand{\s}{\,{\rm s}}      \newcommand{\ps}{\,{\rm s}^{-1}}
    \newcommand{\Msun}{M_{\odot}}   
    \newcommand{\km}{\,{\rm km}}

        \newcommand{\K}{\,{\rm K}}
    
    \newcommand{\G}{\,{\rm G}}

\begin{document}

\title{
Molecular gas distribution perpendicular to the Galactic plane
}

\shorttitle{Vertical Distribution of MCs}

\correspondingauthor{Yang Su}
\email{yangsu@pmo.ac.cn}

\author[0000-0002-0197-470X]{Yang Su}
\affil{Purple Mountain Observatory and Key Laboratory of Radio Astronomy,
Chinese Academy of Sciences, Nanjing 210023, China}

\author{Ji Yang}
\affiliation{Purple Mountain Observatory and Key Laboratory of Radio Astronomy,
Chinese Academy of Sciences, Nanjing 210023, China}
\affiliation{School of Astronomy and Space Science, University of Science and
Technology of China, 96 Jinzhai Road, Hefei 230026, China}

\author{Qing-Zeng Yan}
\affiliation{Purple Mountain Observatory and Key Laboratory of Radio Astronomy,
Chinese Academy of Sciences, Nanjing 210023, China}

\author{Shaobo Zhang}
\affiliation{Purple Mountain Observatory and Key Laboratory of Radio Astronomy,
Chinese Academy of Sciences, Nanjing 210023, China}

\author{Hongchi Wang}
\affiliation{Purple Mountain Observatory and Key Laboratory of Radio Astronomy,
Chinese Academy of Sciences, Nanjing 210023, China}
\affiliation{School of Astronomy and Space Science, University of Science and
Technology of China, 96 Jinzhai Road, Hefei 230026, China}

\author{Yan Sun}
\affiliation{Purple Mountain Observatory and Key Laboratory of Radio Astronomy,
Chinese Academy of Sciences, Nanjing 210023, China}

\author{Zhiwei Chen}
\affiliation{Purple Mountain Observatory and Key Laboratory of Radio Astronomy,
Chinese Academy of Sciences, Nanjing 210023, China}

\author{Chen Wang}
\affiliation{Purple Mountain Observatory and Key Laboratory of Radio Astronomy,
Chinese Academy of Sciences, Nanjing 210023, China}

\author{Xin Zhou}
\affiliation{Purple Mountain Observatory and Key Laboratory of Radio Astronomy,
Chinese Academy of Sciences, Nanjing 210023, China}

\author{Xuepeng Chen}
\affiliation{Purple Mountain Observatory and Key Laboratory of Radio Astronomy,
Chinese Academy of Sciences, Nanjing 210023, China}
\affiliation{School of Astronomy and Space Science, University of Science and
Technology of China, 96 Jinzhai Road, Hefei 230026, China}

\author{Zhibo Jiang}
\affiliation{Purple Mountain Observatory and Key Laboratory of Radio Astronomy,
Chinese Academy of Sciences, Nanjing 210023, China}

\author{Min Wang}
\affiliation{Purple Mountain Observatory and Key Laboratory of Radio Astronomy,
Chinese Academy of Sciences, Nanjing 210023, China}

\begin{abstract}
We use the $\sim$370 square degrees data from the MWISP CO survey to 
study the vertical distribution of the molecular clouds (MCs) toward the 
tangent points in the region of $l=[+16^{\circ},+52^{\circ}]$ 
and $|b|<$5\fdg1.
We find that the molecular disk consists of two components with a
layer thickness (FWHM) of $\sim$~85~pc and $\sim$~280~pc, respectively. 
In the inner Galaxy, the molecular mass in the thin disk is dominant, 
while the molecular mass traced by the discrete MCs with weak CO emission 
in the thick disk is probably $\lsim 10$\% of the whole molecular disk.
For the CO gas in the thick disk, we identified 
1055 high-$z$ MCs that are $\gsim$100~pc from the Galactic plane.
However, only a few samples (i.e., 32 MCs or 3\%) are located in the $|z|\gsim360$~pc region.
Typically, the discrete MCs of the thick-disk population have 
a median peak temperature of 2.1~K, a median velocity
dispersion of 0.8~$\km\ps$, and a median effective radius of 2.5~pc. 
Assuming a constant value of $X_{\rm CO}=2\E{20}$~cm$^{-2}$(K~km~s$^{-1})^{-1}$,
the median surface density of these MCs is $6.8\ \Msun$~pc$^{-2}$,
indicating very faint CO emission for the high-$z$ gas.
The cloud-cloud velocity dispersion is 
4.9$\pm1.3\km\ps$ and a linear variation with a slope of $\sim-0.4\km\ps$kpc$^{-1}$ 
is obtained in the region of $R_{\rm GC}$=2.2--6.4~kpc.
Assuming that these clouds are supported by their turbulent motions 
against the gravitational pull of the disk, a model of 
$\rho_0(R)= 1.28\ \Msun$~pc$^{-3}$~$e^{-\frac{R}{3.2\ {\rm kpc}}}$
can be used to describe the distribution of the total mass density 
in the Galactic midplane. 
\end{abstract}

\keywords{Interstellar medium (847); Molecular clouds (1072); Surveys (1671); 
Milky Way disk (1050); Interstellar dynamics (839); Stellar-interstellar interactions (1576)}

\section{Introduction}
CO is a good tracer of the molecular gas in the interstellar medium (ISM). 
Many CO surveys \citep[e.g., see][]{2015ARA&A..53..583H,2021MNRAS.500.3064S} 
have been undertaken to study the physical properties and the distribution
of molecular clouds (MCs) in the Milky Way.
Astronomers often use these large-scale CO data to investigate
the Galactic structure, e.g., the spiral arms, the molecular gas disk,
the scale height of the gas plane, and some special large-scale 
regions such as the central molecular zone, etc 
(refer to the good summaries in \citealp{1991ARA&A..29..195C}, 
\citealp{2001ApJ...547..792D}, and \citealp{2015ARA&A..53..583H};
also see some recent works based on the large-scale CO data such as 
\citealp{2020A&A...633A.147B,2008ApJ...683L.143D,2011ApJ...734L..24D,
2020MNRAS.498.5936E,2016MNRAS.456.2885R,
2017A&A...601A.124S,2021MNRAS.500.3064S,2019PASJ...71S...1S,2019PASJ...71S...2T}).

The Milky Way Imaging Scroll Painting (MWISP) project is an ongoing large-scale, 
unbiased, and high-quality CO survey toward the northern Galactic plane 
\citep[e.g., see more details in][]{2019ApJS..240....9S}.
The MWISP survey offers an excellent opportunity to improve our understanding 
of the Galactic structure. Based on the rich CO data set of the new survey, actually,
many results on the Galactic structure are also present \citep[e.g., see the different 
arm structures in][]{2016ApJS..224....7D,2017ApJS..229...24D,2016ApJ...828...59S,
2015ApJ...798L..27S,2017ApJS..230...17S,2020ApJS..246....7S}.
Generally, the molecular gas in the inner Galaxy is well confined to the 
plane, while the gas at large Galactocentric distances 
\citep[e.g., $R_{\rm GC}\gsim$~10~kpc in][]{2016ApJ...828...59S} 
often displays the flaring and warping structures.

Additionally, the new CO data reveal that the considerable molecular gas  
is located in the region away from the $b=0^{\circ}$ plane in the inner Galaxy 
because of the larger latitude coverage of $b=[-5$\fdg1, 5\fdg1] in the 
MWISP survey \citep{2019ApJS..240....9S}.
A similar thick molecular disk was investigated both in the Milky Way
\citep[][]{1994ApJ...436L.173D,1994ApJ...437..194M} and other galaxies
(e.g., NGC 891 by \citealp{1992A&A...266...21G} and M51 by \citealp{2013ApJ...779...43P}).
Following the study by \citet{1994ApJ...436L.173D}, we have confirmed that 
the CO disk in the inner Galactic plane is composed of two components, 
a well-known thin disk with an FWHM of 88.5~pc 
and an additional faint thick CO disk with an FWHM of 276.8~pc.
However, the properties and distributions of the molecular gas 
in the Galactic thick disk have not been studied in detail
due to the lack of high-quality large-scale data and large MC samples. 

In this paper, we focus on the vertical distribution of the molecular
gas near the tangent points by using the accumulated CO data from the MWISP. 
Benefiting from the good angular resolution, excellent sensitivity,
and high dynamic range of the MWISP survey, we identified over 1000 MCs 
far from the Galactic plane to trace the thick CO disk in the region of
$l=16^{\circ}$--$52^{\circ}$ and $|b|\lsim5\fdg1$. The properties, 
distributions, and origins of these MCs far from the Galactic plane 
(hereafter the high-$z$ MCs) are investigated accordingly.

The paper is organized as follows. In Section 2, we briefly describe the 
CO observations and data processing. Section 3 shows the results and discussions
on the CO distribution perpendicular to the Galactic disk and 
the MCs far from the midplane. The distribution of the total midplane mass density 
as a function of the Galactocentric distance is also investigated in the section 
based on the estimations of the 
cloud-cloud velocity dispersion and the scale height of the CO gas. 
Finally, we summarize our results in Section 4.

\section{CO Data}
We employed the $\sim370\ {\rm deg}^2$ CO data (i.e., 
$l=16^{\circ}$--$52^{\circ}$ and $|b|\lsim5\fdg1$) from the MWISP project 
\citep[see the details in][]{2019ApJS..240....9S} to study the 
vertical distribution of the molecular gas.
Briefly, the \twCO, \thCO, and C$^{18}$O ($J$=1--0) lines were
simultaneously observed with the full-sampling On-The-Fly mapping
\citep[see][]{2018AcASn..59....3S} by using the nine-beam 
Superconducting Spectroscopic Array Receiver system \citep{Shan}. 
The spatial and spectral resolutions of the CO data are $\sim50''$ 
and $\sim$~0.2~$\km\ps$, respectively. The quality of the CO data is good,
and the first-order (or linear) baseline was fit for all spectra. 
After removing the bad channels and abnormal spectra, the reduced 3D data 
cubes (position-position-velocity, hereafter PPV) with a uniform grid 
spacing of 30$''$ have typical rms noise levels of 
$\sim$~0.5~K for \twCO\ at a channel width of 0.16~$\km\ps$ and 
$\sim$~0.3~K for \thCO\ (C$^{18}$O) at 0.17~$\km\ps$, respectively.

\section{Results and Discussions}

\subsection{Terminal Velocities from the MWISP CO Data}
Figure~\ref{guidemap} displays the schematic view of the
molecular gas discussed in this paper. 
The red belt with a length of $\sim$5.1~kpc shows the region of the tangent-point MCs
between $l=16^{\circ}$ and $l=52^{\circ}$. The blue circle indicates
the approximate radius of the end of the bar (i.e., the 3 kpc ring structure).
The gas emission inside of the circle drops sharply.
Assuming a circular motion of MCs in the Milky Way,
we can easily obtain the distances ($d=R_{0}\times{\rm cos}\ l$)
and Galactocentric distances ($R_{\rm GC}=R_{0}\times{\rm sin}\ l$)
for the MCs at the terminal velocity.
Here, the value of the Sun's distance to the Galactic center, $R_{0}$, 
is taken to be 8.15~kpc \citep[see the recent work by][]{Reid19}.

The CO emission near the tangent points is mainly concentrated in
the region of $|b|\lsim$1\fdg5 \citep[e.g., see Figure 6 in][]{2019ApJS..240....9S}.
Considering the large-scale distribution of the CO gas near the tangent points,
we divide the whole region into 36 subregions to determine the corresponding 
terminal velocity in the range of $l=16^{\circ}-52^{\circ}$. 
Each subregion is centered at b=0~$^{\circ}$ and has an area of 
3~deg$^2$, i.e., $\Delta l=1^{\circ}$ and $b=[-1$\fdg5, 1\fdg5].
The average spectra of these subregions were extracted to search
for the CO peak emission at the most positive velocity, i.e., 
the tangent velocity, $V_{\rm tan}(l)$, for each sampled Galactic longitude. 
At the tangent point, the typical value of the average peak \twCO\ temperature 
of each subregion is selected to be $T_{\rm peak12}\gsim$~0.5~K for $l=22^{\circ}-52^{\circ}$ 
and $T_{\rm peak12}\gsim$~0.2~K for $l=16^{\circ}-22^{\circ}$. 
That is, the smaller features (and thus the MCs with very weak CO emission) 
are ignored because we only focus on the ensemble of the large-scale 
molecular gas near the tangent points in each subregion.
These smaller features with somewhat larger $V_{\rm LSR}$ are probably 
due to local variations of the environment (i.e., the abnormal velocities
from star-forming activities, e.g., \HII\ regions, star winds, supernova remnants,
etc.) and noncircular motions from other effects near the tangent points.
For blended and confused CO emission, the combined \thCO\ line 
is also used to define the CO peak emission at the most positive velocity 
(i.e., $V_{\rm tan}(l)$ at $T_{\rm peak13}\gsim$~0.2~K).

The above approach to define the terminal velocity can avoid the effects of 
the contaminated emission from the wing of a component and noncircular motions
from some smaller molecular features (e.g., the redshift velocity from 
the perturbed gas near the tangent points).
We plot the determined curve of the terminal velocity on the longitude-velocity 
diagram (i.e., the LV map) by interpolating for the MWISP CO data 
in the region of $l=[16^{\circ}, 52^{\circ}]$ (see the red line in Figure~\ref{pv12}).
We find that our result is consistent with previous CO studies
\citep[e.g., see][]{1985ApJ...295..422C,Malhotra94}.

For further comparisons of the tangent-point features at the most positive velocity, 
we also overlaid the terminal velocity curves from 
the \HI\ observations \citep[the gold lines;][]{2016ApJ...831..124M} 
and the trigonometric parallaxes of high-mass star-forming regions 
\citep[the purple line; see the A5 model from][]{Reid19}
on the CO LV map, respectively.
Generally, for the discussed region, the $l$--$V_{\rm tan}$ trend is 
very similar between our CO result and the \HI\ result, although
our CO terminal velocity is systematically smaller (several $\km\ps$)
than that of the \HI\ observations. 
Actually, we find that the \HI\ terminal-velocity curve is 
nearly around the outer layer of the CO emission at the most extreme velocity
(see the thick golden line in Figure~\ref{pv12}). 
The difference between the CO result and the \HI\ result is probably due to
(1) the different properties of the two tracers,
and (2) the different methods used in studies.

On the other hand, models based on different tracers also display
somewhat differences in calculating the terminal velocity
at different Galactic longitudes (e.g., the thin golden line vs. 
the purple line in Figure~\ref{pv12}). 
Further investigations are helpful to clarify these issues 
by combining the \HI\ and CO data in a larger longitude range.
In this paper, we tentatively use the MWISP CO observations to trace
the terminal velocity 
(see the red line in Figure~\ref{pv12}) for the subsequent analysis.
As discussed below, the slight variation of the terminal velocity
does not change our analysis and result considerably.

\subsection{Thin and Thick Molecular Disks}
Using the terminal velocity curve determined in Section 3.1,
we can investigate the vertical distribution of the molecular gas
at the tangent points. We made the integrated \twCO\ emission
in the velocity range of $V(l)\gsim V_{\rm tan}(l)-7$~km~s$^{-1}$ 
to trace the molecular gas at the most positive velocity (Figure~\ref{tan30cut}).
Some large-scale structures near the tangent points are labeled
on the map \citep[e.g., refer to][]{2016ARA&A..54..529B,2016ApJ...823...77R}.
The value of 7~km~s$^{-1}$ was adopted to take into account 
the cloud-cloud velocity dispersion of the molecular gas 
near the tangent points \citep[e.g., see][]{Malhotra94}.
Actually, this value is roughly consistent with the 
maximum cloud-cloud velocity dispersion measured from the tangent-point MCs
based on the MWISP CO observations (Section 3.4).

Based on the \twCO\ emission near the most positive velocity
(Figure~\ref{tan30cut}), we find that molecular gas is
mainly concentrated around the plane in the inner Galaxy
(e.g., $|b|\lsim1^{\circ}$ or $|z|\lsim$100~pc).
The whole region from $l=[+16^{\circ},+52^{\circ}]$ is divided into 
six ranges of longitude to investigate the vertical distribution of the molecular gas.
For a certain region, CO emission at the same distance from the 
Galactic plane of $b=0^{\circ}$ (i.e., $z=R_{\rm GC}$cos($l$)tan($b$))
is integrated assuming the gas is located at the tangent points.
Along the direction perpendicular to the Galactic plane,
the total intensity of the CO emission with a bin of 5~pc 
is then fitted by using a Gaussian function, 
i.e., $I(z)/I_{\rm peak}$=exp$(-\frac{(z-z_0)^2}{2{\sigma_z^2}})$.
Here, $z_0$ and $\sigma_z$ (i.e., the thickness FWHM=2.355$\sigma_z$) 
in units of parsec is the mean value and the standard deviation of
the vertical distribution of the CO emission, respectively. 
Note that the CO gas outside the $|z|\gsim$100--150~pc region
does not affect our single-Gaussian fitting because the high-$z$ emission
is very weak in the total CO intensity near the tangent points.

The best fit of the vertical distribution of the \twCO\ emission is shown 
in Figure~\ref{narrow}, while the result of the \thCO\ emission is also 
shown for comparison.
Obviously, the $z_0$ values from the \twCO\ emission agree well with 
those from the \thCO\ emission, indicating that the molecular gas 
traced by the optically thick and thin lines is concentrated 
toward the midplane.
We find that the mean thickness of the \twCO\ disk is roughly 94~pc
between $l=22^{\circ}$ and $l=52^{\circ}$, which is larger than
the value of $\sim$60~pc in the region of $l=16^{\circ}-22^{\circ}$.
On the whole, the narrow Gaussian component represents well
the thin molecular gas disk that is dominated by
the enhanced CO emission near the Galactic midplane.

Meanwhile, there are amounts of CO gas located far from the
Galactic plane (Figure~\ref{tan30cut}). 
To investigate the distribution of the gas far from the plane, 
we use the mean intensity in each 5~pc bin to reveal the weak CO emission.
That is, we use the mean intensity in an effective area 
(i.e., $I_{\rm total}/\sqrt{N_{\rm pixel}}$ for pixels of $I_{\rm CO}\gsim 3I_{\rm rms}$) 
to investigate the distribution of the CO gas in the disk. Obviously, 
this processing nearly does not affect the distribution of the gas in the thin disk
because of the strong CO emission there. However, it can highlight the
distribution of the weak CO emission in the high-$z$ region because
only pixels with $I_{\rm CO}\gsim 3I_{\rm rms}$ are accounted for.
By adopting the known values of $z_0$ and $\sigma_z$ for the thin CO disk
(i.e., the narrow Gaussian component from Figure~\ref{narrow}), 
we can easily fit another broad Gaussian component for the thick CO disk 
(Figure~\ref{broad}).

As a result, a mean FWHM of $\sim$256~pc (excluding the region of 
$l=16^{\circ}-22^{\circ}$) can well describe the thick molecular gas disk. 
It is worth mentioning that the gas distribution of the thick disk is not 
symmetric with respect to the thin CO disk in some regions.
Then the fitted $z_0$ values of the thick disk are not always consistent with 
those from the thin CO disk (e.g., see the enhanced CO emission at $z\lsim-100$~pc 
in the longitude range of $22^{\circ}-28^{\circ}$ in Figure~\ref{broad}). 
Actually, we also note that there are lots of atomic clouds below the Galactic
plane in a similar region (e.g., see the $l=22^{\circ}-31^{\circ}$ and
$b\lsim-2^{\circ}$ region in Figure~\ref{CO_HI}).
This feature probably indicates that the gas of the thick disk is probably decoupled 
from the thin gas layer in some regions.

Similarly, we also measured the mean thickness of the molecular disk in the whole range of 
$l=16^{\circ}-52^{\circ}$ (the upper panels in Figure~\ref{disk1652}) and 
$l=22^{\circ}-52^{\circ}$ (the lower panels in Figure~\ref{disk1652}) 
by using the  best fit of the two Gaussian components.
The thickness of the thin and thick CO disk is roughly 107~pc and 301~pc, respectively. 
Both of the two values from the wider longitude region are 
$\sim$15\% larger than the results from the smaller subregions of $\Delta l=6^{\circ}$ 
(i.e., 107~pc vs. 94~pc for the thin CO disk and 301~pc vs. 256~pc for the thick CO disk).

We further investigate the variation of the thin disk 
over the Galactic longitude. Based on the measurements 
of the 30 subregions with $\Delta l=1^{\circ}$, 
the mean thickness of the thin CO layer (excluding $l=16-22^{\circ}$) 
is $\sim$85~pc (see the right panel in Figure~\ref{z0width}), 
which is slightly smaller than values of $\sim$94--107~pc from
Figures~\ref{narrow} and \ref{disk1652}.
We note that the mean thickness of the thin CO disk remains almost
unchanged in the region of $l\sim22-52^{\circ}$, while the thickness of the
$l\sim16-22^{\circ}$ region is considerably smaller (see the right panel in Figure~\ref{z0width}).
According to Figure~\ref{tan30cut}, we also note that the CO emission
in $l=16^{\circ}$ and $l=22^{\circ}$ is relatively weak with respect to other regions
near the tangent points. We speculated that it is probably due to the effect
of the dynamical perturbation within the 3 kpc ring.

The larger values of the thickness of the CO layer from Figure~\ref{disk1652}
can be attributed to the variation of $z_0$ at different Galactic 
longitudes, which will widen the thickness of the gas disk by fitting data from 
the larger longitude range. Indeed, we find that $z_0$ changes from negative values
at $l\sim37-52^{\circ}$ to positive values at $l\sim16-37^{\circ}$
(see the left panel in Figure~\ref{z0width}).
As discussed in \citet{2015ARA&A..53..583H}, other methods 
probably yield a higher value 
for the CO layer thickness due to corrugations in the inner Galactic disk
\citep[e.g., FWHM$\sim$100--120 pc in][]{1988ApJ...324..248B,1988ApJ...327..139C,
1984ApJ...276..182S,2006PASJ...58..847N}.

We also check our fitting by using the integrated CO emission from
somewhat different velocity ranges, e.g.,
$V(l)\gsim V_{\rm tan}(l)-4$~km~s$^{-1}$ and
$V(l)\gsim V_{\rm tan}(l)-10$~km~s$^{-1}$.
The resultant fittings are not changed significantly.
To recapitulate briefly, our recommended thickness of the thin CO layer 
of $\sim$85~pc is in agreement with the value of $\sim$90~pc from 
\citet{Malhotra94}, and the thickness of the thick CO layer 
is about 260--300~pc, which is also consistent with previous studies 
\citep[e.g.,][]{1994ApJ...436L.173D}.
The thick molecular disk revealed by the faint CO emission is about three times 
as wide as the well-known central thin CO layer. The thickness of the thick CO
layer is also comparable in width to the central \HI\ 
layer \citep[e.g., 250-270~pc in][]{Dickey90,Lockman91}.

Moreover, the thickness of the thick CO layer from the MWISP survey is just 
between $\sim190$~pc for the diffuse H$_2$ gas and $\sim320$~pc for the 
diffuse \HI+warm ionized medium (WIM) gas by considering $R_0=$8.15~kpc 
\citep[see the discussions of the bright/faint diffuse \CII-without-CO emission in]
[]{Velusamy14}.
These results show that a considerable amount of molecular gas in the  
CO-faint H$_2$ clouds can be revealed by the high-resolution and 
high-sensitivity CO survey.
In the following section, we will investigate the distributions and 
properties of the CO-faint molecular gas based on the newly identified 
high-$z$ MCs near the tangent points.

\subsection{MCs Far from the Galactic Plane}

\subsubsection{Identification of the High-$z$ MCs}
MCs far from the Galactic plane are identified by using the 
density-based spatial clustering of applications with noise 
(DBSCAN\footnote{https://scikit-learn.org/stable/auto\_examples/cluster/plot\_dbscan.html}) 
clustering algorithm.
Full details can be found in \citet{2020ApJ...898...80Y},
and a brief description of the method is presented below. 

First, the 3D PPV data cubes in the velocity
interval of 40--160$\km\ps$ are smoothed to 
0.5$\km\ps$ to decrease the random noise fluctuations in the velocity axis.  
In the PPV space, the minimum number of neighborhood voxels (MinPts)
is set to 16 and the minimum cutoff on the data is 2$\times$rms
(i.e., the lowest level to surround the 3D structure). 
The peak brightness temperature ($T_{\rm peak}$) is $\geq 5\times$rms 
(here, the rms is $\sim$0.3~K for the smoothed channel width of 0.5$\km\ps$
for \twCO)
to control the quality of the selected samples. 
Second, the projection area on the spatial scale has at least 4 pixels 
($\sim$ one beam), and the minimum channel number in the velocity axis 
is considered to be $\geq$3 to reduce striping artifacts
and other uncertainties in the whole data.
These criteria are helpful to obtain as much of the emission 
as possible and to avoid the contamination of the noise in the 
3D data, leading to weak but true signals for the large-scale CO data
to be picked up.
Broad criteria (e.g., MinPts=12 and/or $T_{\rm peak}=3\times$rms) 
will increase the number of the MC sample; however, there is usually
a large uncertainty for the true MC identification due to the 
confusion with the noise fluctuations.

Finally, MCs far from the Galactic plane are selected based on
the further criteria of $V_{\rm MC}\gsim V_{\rm tan}(l)-7$~km~s$^{-1}$
and $z_{\rm MC}(l,b,v) \leq z_0(l) -3\sigma_z(l)$ or
$z_{\rm MC}(l,b,v) \geq z_0(l) +3\sigma_z(l)$. 
Here, $z_0(l)$ and $\sigma_z(l)$ can be obtained based on the best fit
of the $z_0(l)$--$l$ and $\sigma_z(l)$--$l$ relations from the 
thin CO disk (see the measurements and the fitting lines in Figure~\ref{z0width}). 
That is, MCs in the thin CO disk are excluded due to their 
concentrated distribution in the region of $b\sim0^{\circ}$,
i.e., $z_0(l) -3\sigma_z(l) < z_{\rm MC} (\rm{thin\ disk})$ $< z_0(l) +3\sigma_z(l)$.

In total, 1055 samples near the tangent points were identified as 
the MCs far from the Galactic plane between $l=16^{\circ}$ and $l=52^{\circ}$ 
based on the MWISP \twCO\ data.
The MCs, which have weak CO emission, are relatively isolated with respect to 
the molecular gas near the Galactic midplane. 
Table~1 lists the parameters of each MC, i.e, 
(1) the ID of the identified MCs, arranged from the low Galactic longitude; 
(2) and (3) the MC's Galactic coordinates ($l$ and $b$); 
(4) the MC's LSR velocity ($V_{\rm LSR}$); 
(5) the MC's one-dimensional velocity dispersion ($\sigma_v$); 
(6) the MC's peak emission ($T_{\rm peak}$); 
(7) the MC's area; 
(8) the MC's distance obtained from the tangent points, i.e, $d=R_0\times$cos($l$);
(9) the MC's $z$ scale defined as $z=d\times$tan($b$);
(10) the MC's mass estimated from the CO-to-H$_2$ conversion factor method, 
$X_{\rm CO}=2\E{20}$~cm$^{-2}$(K~km~s$^{-1})^{-1}$ \citep[e.g.,]
[]{2001ApJ...547..792D,2013ARA&A..51..207B}; 
and (11) the MC's virial parameter $\alpha=\frac{M_{\rm virial}}{M_{\rm Xfactor}}
=\frac{5\sigma_{v}^2R}{GM_{\rm Xfactor}}$,
where $R$ is the radius of the MC and $G$ the gravitational constant.
We can easily use MWISP Glll.lll$\pm$bb.bbb$\pm$vvv.vv 
to name an MC identified from the CO survey.


\subsubsection{Properties and Statistics of the High-$z$ MCs}
For the molecular gas far from the Galactic plane, Figure~\ref{sta_zmass}
displays the vertical distribution of the 1055 MCs, which can be fitted by
a Gaussian function with $z_0=-15.9$~pc and $\sigma_z=113.1$~pc.
The standard deviation of the vertical distribution is $\sim$113~pc 
(and thus the FWHM of $\sim$270~pc) for the gas layer traced by 
the ensemble of the high-$z$ MCs, 
which is roughly similar to the measurement from the 
CO emission in the thick molecular gas layer (Section 3.1).

Importantly, the excellent agreement of the vertical distribution of these
discrete MCs with that of the thick CO disk emission proposed by \citet{1994ApJ...436L.173D}
confirms their claim that the high-$z$ emission they observed is not substantially 
contaminated by sidelobe pickup from the central disk.
Therefore, our results based on the large-scale, high-resolution, and high-sensitivity 
data demonstrate convincingly that the thick molecular gas disk is indeed 
an important component in the inner Galaxy. The thick molecular disk is 
composed of many discrete MCs with small size and low mass (see below).

Figure~\ref{sta_para} shows the properties of the MCs.
Typically, these MCs have $T_{\rm peak}\sim$1--4~K, 
with the median value of 2.1~K, indicating very faint CO emission 
(i.e., $\sim$4--10~$\Msun$pc$^{-2}$ assuming a constant CO-to-H$_2$ 
conversion factor) for these small MCs. 
The effective radii of these MCs are 0\farcm8--1\farcm6 
($\sim$1--2 beam size), indicating that the MCs at the tangent are 
unresolved because of the spatial resolution limit of the data, 
i.e., 1--4 pc at distances of 5--8~kpc.
We find that a truncated power-law function with a slope of 
$\gamma$=-1.74 (i.e., $N(M>M_0) \propto M^{\gamma+1}$ for the cloud's mass range
of $\sim120-8000\ \Msun$) can describe the mass distribution of the high-$z$ MCs,
which is consistent to $\gamma \sim$-1.7 observed for the MCs in 
the Milky Way \citep[e.g.,][]{2010ApJ...723..492R,2015ARA&A..53..583H,
2016ApJ...822...52R,2019MNRAS.483.4291C}.
The velocity dispersion ($\sigma_v$) of the high-$z$ MCs is roughly 0.8$\km\ps$,
leading to $\alpha\gsim10$ for the most of CO clouds.

Obviously, the lower limit of $T_{\rm peak}$, radius, and surface
density is related to (1) the limited sensitivity and spatial resolution of 
the current CO survey, and (2) the selection criterion of the MCs used in Section 3.3.1.
We mention that some clouds with weak CO emission
($T_{\rm peak}\lsim 1$~K) and/or smaller sizes (radius$\lsim 1'$)
are missed due to the observational limit of the MWISP survey.
Therefore, a substantial amount of the fainter CO emission 
(e.g., $\lsim1$~K~km~s$^{-1}$ or $M \lsim 100-150\ \Msun$) 
could probably be unveiled by higher-sensitivity CO surveys.

The cloud-cloud velocity dispersion ($\sigma_{\rm cc}$) of these 
high-$z$ MCs can be estimated from the samples with $V_{\rm LSR}\gsim V_{\rm tan}$.
Obviously, the estimation of $\sigma_{\rm cc}$ depends on the definition of the $l-V_{\rm tan}$ 
relation. By taking into account the slight variation of the tangent velocity, 
we find that the cloud-cloud dispersion varies from $4.4\pm1.6\km\ps$
for $V_{\rm LSR}\gsim V_{\rm tan}$
to $5.6\pm1.5\km\ps$ for $V_{\rm LSR}\gsim V_{\rm tan} - 7\km\ps$
with the $1^{\circ}$ bin samples in the $l=[+16^{\circ}, +52^{\circ}]$ region.
The cloud-cloud velocity dispersion only changes a little with the
variation of the selected tangent velocity.
On the other hand, the estimated $\sigma_{\rm cc}$ of these MCs is     
comparable to the measured cloud-cloud velocity dispersion of the total MC samples 
near the Galactic plane \citep[see Section 3.4 and other works, e.g.,][]
{1989ApJ...339..763S,Malhotra94}.

Interestingly, the low $\sigma_{\rm cc}$ of the high-$z$ MCs in the Milky Way 
is different from the finding that a potential thick gas disk may have 
a high velocity dispersion for galaxies \citep[e.g., $\sigma_{\rm cc} \sim 12\km\ps$ 
for the thick molecular gas disk with extended/diffuse CO emission;][]{2013AJ....146..150C}.
As discussed in \citet{2018MNRAS.477.2716K}, the high mass transport rates
and star formation rates can lead to the high velocity dispersions 
of the gas in galaxies. 
Briefly, we summarized all of the physical properties for the high-$z$ MCs in Table~2.

\begin{figure}
\includegraphics[trim=0mm 0mm 0mm 0mm,scale=0.9,angle=0]{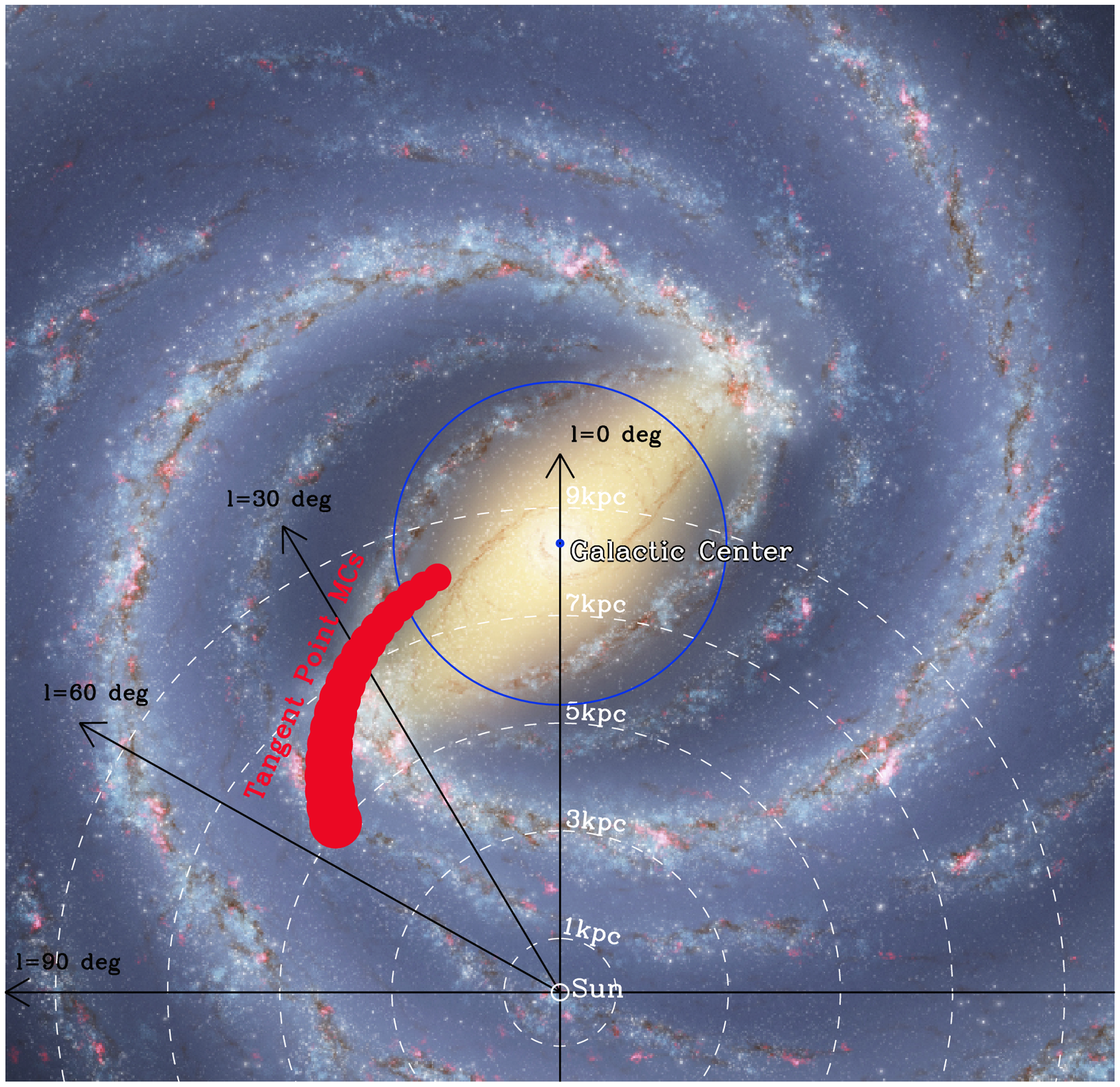}
\caption{
Face-on view of the Milky Way (R. Hurt: NASA/JPL-Caltech/SSC) superposed on
the tangent-point molecular gas of the Galaxy.
The MCs near the tangent points are identified based on the MWISP CO data
toward the longitude range of 16$^{\circ}$--52$^{\circ}$ 
(or a red belt with a length of $\sim$5.1~kpc). Note that the
width of the belt in the map is proportional to the longitude, and 
the mean width is about 500~pc based on the model \citep[e.g.,][]{1984ApJ...283...90L}.
The Sun at [0, -8.15]~kpc and the Galactic center at [0, 0]~kpc are also 
labeled on the map. The blue circle indicates the 3 kpc ring with 
respect to the Galactic center.
\label{guidemap}}
\end{figure}

As seen in Figure~\ref{sta_z}, the MCs are mainly concentrated in regions
of $l=22^{\circ}-30^{\circ}$ (i.e., $R_{\rm GC}$=3.1--4.1~kpc) 
and $l=43^{\circ}-52^{\circ}$ (i.e., $R_{\rm GC}$=5.6--6.4~kpc).
A smaller concentration is between the two peak distributions
(i.e., see the $l=32^{\circ}-38^{\circ}$ or $R_{\rm GC}$=4.3--5.0~kpc 
region in Figure~\ref{sta_z}).
We suggest that the large-scale structures (i.e., the Norma arm, 
the Aquila Spur toward the tip of the Galactic bar,
and the Carina-Sagittarius arm; see Figures~\ref{guidemap} 
and \ref{tan30cut}; refer to \citealp{2016ARA&A..54..529B,2016ApJ...823...77R}) 
near the tangent points are probably responsible for 
the three peaks at the longitude distribution (or $R_{\rm GC}$) of the enhanced high-$z$ MCs.
Interestingly, the scatter of $\sigma_{\rm cc}$ is relatively larger in the 
region of $l\sim22^{\circ}-38^{\circ}$ (Figure~\ref{sigv_12}), 
which indicates that the Galactic bar may have an important impact
on the dynamics and distribution of the gas in the inner Galaxy.

We also find that the high-$z$ \HI\ gas is abundant in 
the region of $l\sim20^{\circ}-30^{\circ}$ \citep[][]{Ford10}, 
although the vertical distribution of the high-$z$ \HI\ clouds is much wider
than that of the CO clouds
(i.e., 300~pc$\lsim |z_{\rm {HI\ cloud}}|\lsim 1700$~pc vs. 
100~pc$\lsim |z_{\rm {CO\ cloud}}|\lsim 250$~pc in the 90\% range).
Figure~\ref{CO_HI} shows the comparison between the high-$z$ MWISP CO clouds
and the GASS \HI\ clouds in the range of l=[+17\fdg1,+34\fdg2]
and b=[+5\fdg1,$-$5\fdg1].
We note that some \HI\ samples in \citet[][]{Ford10}
are very likely related to our CO clouds, e.g., 
(1) the GASS \HI\ cloud: (18\fdg72, $-$2\fdg00, 126.1~$\km\ps$) versus
the MWISP CO cloud: (18\fdg805, $-$1\fdg937, 131.27~$\km\ps$);
(2) the GASS \HI\ cloud: (19\fdg19, $+$2\fdg06, 136.3~$\km\ps$) versus
the MWISP CO cloud: (19\fdg213, $+$2\fdg134, 137.38~$\km\ps$);
(3) the GASS \HI\ cloud: (19\fdg61, $+$4\fdg43, 138.0~$\km\ps$) versus
the MWISP CO cloud: (19\fdg512, $+$4\fdg377, 137.84~$\km\ps$), etc.
The spatial and kinematical relationships between the \HI\ cloud
and the CO cloud are worth exploring in more detail in future works 
by using the high-quality \HI\ data \citep[e.g., from the GALFA-HI survey;][]
{2011ApJS..194...20P,2018ApJS..234....2P} and MWISP CO data. 
Here we briefly discuss the possible origin of the high-$z$ MCs in 
Section 3.3.3.

In the region of $|z-z_0|>\sigma_z$ (i.e., the region outside the $1\times\sigma_z$ of
the thick CO disk), the molecular gas mass is about $3.6\times10^5\ \Msun$ based on
the discovered high-$z$ MCs by adopting $z_0$=-15~pc and $\sigma_z$=120~pc 
(e.g., see Figure~\ref{sta_zmass}). 
The value is approximately $31.7\%$ of the total mass of the thick CO disk 
in the 5.1~kpc long red belt (see Figure~\ref{guidemap}; including an extrapolation 
of the component through the full Galactic disk by assuming a Gaussian distribution). 
Thus, the total mass of the thick CO disk in the belt region is about 
$1.1\times10^6\ \Msun$. Adopting a mean width of 0.5~kpc for the red belt 
\citep[i.e., the mean path length along the line of sight for the belt; 
see the model in][]{1984ApJ...283...90L},
we obtain an area of 2.55~kpc$^2$ (i.e., 5.1~kpc length $\times$~0.5~kpc width) 
projected on the Galactic plane. The total midplane density of the thick disk is thus 
$\sim0.43\ \Msun {\rm pc}^{-2}$, leading to the volume density of 
$\sim0.0014\ \Msun {\rm pc}^{-3}$ (or $\sim0.02\ {\rm H_2}{\rm cm}^{-3}$) in the
discussed region by assuming the Gaussian distribution of 
$\rho(z)=\rho_0$exp$(-\frac{(z-z_0)^2}{2{\sigma_z^2}})$.
We mentioned that the estimated values are probably the lower limit due to 
the limited observational sensitivity and resolution of the MWISP survey
(i.e., the missing molecular gas mass for the very faint CO emission and small clouds
that are not involved in the calculation).
In the region of $R_{\rm GC}=$2--8.15~kpc, the molecular mass 
of the thick CO disk is thus $\sim8.5\times10^7\ \Msun$, which is about 10\% of
the total molecular mass in the inner Galaxy \citep[i.e., $\sim7.8\times10^8\ \Msun$ 
for the thin CO disk + the thick CO disk; this paper and][]{2015ARA&A..53..583H}.

\subsubsection{Origin of the high-$z$ MCs}
According to the vertical distribution of the high-$z$ molecular gas (Figure~\ref{sta_zmass}), 
we find that the identified MCs are mainly concentrated in the 100~pc$\lsim |z| \lsim$~360~pc region.
However, there is only a little amount of molecular gas traced by CO emission in the 
region of $|z|\gsim$~360~pc (i.e., 32/1055 high-$z$ MCs). 
We thus suggest that these MCs with faint CO emission 
probably belong to the disk population.
The spatial and velocity coincidence between some CO clouds 
and the corresponding \HI\ clouds is very interesting. 
It indicates that considerable amounts of molecular gas 
appear to survive in the disk-halo transition region (or in the 
disk-halo zone close to the Galactic midplane; e.g., 100~pc$\lsim |z|\lsim$~300~pc).
The coexistence of molecular gas and the atomic gas in
the disk-halo transition region needs to be further investigated
based on the observational and theoretical studies.

Generally, the typical internal crossing time of these high-$z$ MCs 
is about 3~Myr assuming the typical radius of $\sim$3~pc and velocity dispersion 
of $\sigma_v=0.9 \km\ps$ (Figure~\ref{sta_para}). This value is roughly 
an order of magnitude smaller than
the dynamical timescale of the moving clouds from the Galactic midplane
(i.e., t$_{\rm dyn}=\frac{z}{\sigma_{\rm cc}}\gsim$24~Myr,
where $\sigma_{\rm cc}=4.9\km\ps$ is the cloud-cloud velocity dispersion
from Section 3.4). It indicates that the MCs' memory of their birthplace
is lost, and any observed kinematical features from the line profile and
spatial morphology probably represent the turbulence in the current local environment.
Considering the prevalent turbulence in the ISM, the cloud will be destroyed
quickly due to the Kelvin-Helmholtz and Rayleigh-Taylor instabilities 
\citep[e.g, $\lsim 1$~Myr in][]{2019ApJ...873....6I}.

On the other hand, the high virial parameter of the clouds shows that the high-$z$ MCs 
are unstable unless they are confined by some external pressure. The low $\sigma_{\rm cc}$ 
of $\sim5\km\ps$ and large $\sigma_z$ of 260--300~pc for these high-$z$ MCs 
indicate that the gravitational force of the midplane is not balanced by the 
gas pressure, suggesting that the ensemble of the MCs is in general out of equilibrium. 
All of these results show that the high-$z$ MCs probably have short
lifetimes less than several Myr (e.g., comparable to the typical internal crossing time).
Indeed, if the MCs directly move from the Galactic plane to their current places,
the moving velocity of the clouds should be an order of magnitude larger than
that of the cloud-cloud dispersion velocity because of the short lifetime of
these clouds (e.g., $\lsim$several Myr).
Alternatively, some of the high-$z$ MCs are probably newborn clouds
due to the rapid H$_2$ formation in shock interaction regions
\citep[e.g., with the high ram pressure of $\gsim10^5-10^6\ {\rm K\ cm^{-3}}$,
see][]{2018ApJ...863..103S}. We suggest that these MCs are transitory.
The high-$z$ molecular gas with faint CO emission is either dispersing
or being assembled by some external dynamical processes
(e.g., compression by shocks, cloud-cloud collision, and shear motions, etc).

Whatever the exact formation mechanism of the high-$z$ clouds, 
the energetic sources near the Galactic plane may play important roles 
in the origin of the disk-population gas
\citep[e.g., the stellar feedback from massive stellar winds and/or supernova explosions;
see the Galactic fountain models in][]{1976ApJ...205..762S,
1980ApJ...236..577B,1990ApJ...352..506H,2008A&A...484..743S,2020A&A...642A.163S}.
Recently, \citet{2020Natur.584..364D} also found that the cold, dense, and
high-velocity molecular gas survives in the Milky Way's nuclear wind
at $\sim$600--900~pc from the Galactic plane.
Likewise, the interactions between the disk gas and the halo gas are 
probably common in the Milky Way and other galaxies due to energetic processes 
in the galactic plane \citep[e.g., see the disk-halo scenario discussed in][]
{2002ApJ...580L..47L,2008ApJ...688..290F,Ford10,2012ARA&A..50..491P}.

Finally, we summarize the thickness of the inner disk revealed by the different gas
tracers in Table~3. 
The fact that the thickness of the thick disk traced by the CO gas
($\sim$260--300~pc in this work) and the \HI(\CII) gas
\citep[e.g., $\sim$250--270~pc for \HI\ emission and $\sim$190--320~pc for \CII\ emission,]
[]{Dickey90,Lockman91,Langer14,Velusamy14}
is comparable may be a relevant hint to establishing a link between 
the formation/evolution of the high-$z$ MCs and various physical processes.
Further studies will be helpful to clarify these issues through a combination 
of multiwavelength data, e.g., the kinematical connection between the CO clouds
and the \HI\ clouds, possible related IR features, and/or emission from other 
tracers like OH 18~cm, CH 3.3~GHz lines, and 158~$\mu$m [\CII] lines, etc.




\subsection{Cloud-to-cloud Velocity Dispersion near the Tangent Points 
and the Total Midplane Density}
The \HI\ and H$_2$ gas in the inner Galaxy ($R_{\rm GC}=R_{0}\lsim8.15$~kpc) 
is concentrated toward the thin plane due to the gravitational potential of 
the matter in the Galactic disk.
If the pressure from magnetic fields, cosmic rays, and the radiation field is ignored,
the vertical distribution of the H$_2$ gas is mainly determined by the total
gravitating mass near the disk (i.e., stars, gas, and possible other unknown objects).
Considering the balance between the turbulent pressure of the 
isothermal molecular gas and the gravitational force on the Galactic plane,
we can obtain the formula of $\rho_0=\frac{\sigma_{\rm cc}^2}{4 \pi G {\sigma_z}^2}$
for the Gaussian distribution of the gas layer 
\citep[e.g., see][]{Malhotra94,1995ApJ...448..138M}.
Here, $\rho_0$ is the total density of the midplane mass,  
$\sigma_{\rm cc}$ is the cloud-cloud velocity dispersion,
and $\sigma_z$ is the scale height of the thin CO layer.
We ingore the scale height of the thick CO disk due to 
its limited contribution to the total molecular gas mass 
(i.e., $\sim$3\% for the $|z|\gsim$120~pc molecular gas).


In principle, we can estimate the midplane mass density of the Galactic disk
if $\sigma_{\rm cc}$ and $\sigma_z$ are obtained from observations.
Figure~\ref{sigv_12} displays the distribution of $\sigma_{\rm cc}$ for all MCs
near the tangent points (i.e., MCs in the thin and thick disk with
$V_{\rm LSR}(\rm {MC})\gsim V_{\rm tan}-\Delta V$).
Here, $\Delta V$ is simply selected as $3.9 \km\ps$ (i.e., roughly three times of
the scatter of $\sigma_{\rm cc}$, which is nearly constant for the tangent MCs
in the longitude range) by taking into account the tangent MCs with somewhat 
lower $V_{\rm LSR}$. 
We note that the estimated value of $\sigma_{\rm cc}$ is in the range of 
2.7--7.9$\km\ps$ (i.e., mean value of 4.9$\km\ps$ and a scatter of 1.3$\km\ps$), 
which is close to that of the high-$z$ MCs discussed in Section 3.3.2.
The value is roughly consistent with other works 
based on different approaches \citep[e.g., see][]{1984ApJ...282L...9B,
1985ApJ...295..422C,1984ApJ...281..624S,1989ApJ...339..763S,Malhotra94}.

Additionally, the distribution of $\sigma_{\rm cc}$ seems to display a linear variation
with a slope of $\sim-0.4\km\ps$kpc$^{-1}$ between $R_{\rm GC}$=2.2--6.4~kpc.
It is interesting to note that the slope from the CO data is about one-half of 
the \HI\ gas when the joint gravitational potential is considered
\citep[i.e., stars, \HI\ gas, and H$_2$ gas near the disk; see][]{2002A&A...394...89N}.
The decrease of $\sigma_{\rm cc}$ at larger $R_{\rm GC}$ may be explained 
by a lower star formation rate in the outer part of the inner disk
if $\sigma_{\rm cc}$ (i.e., the turbulence of the MCs) is regulated by the 
energy input via star-forming activities near the Galactic plane.
On the other hand, however, it should be mentioned that the measured 
$\sigma_{\rm cc}$ based on our CO data also displays considerable fluctuation 
in different $R_{\rm GC}$ (see Figure~\ref{sigv_12}), which is probably related to 
the different localized environment. 
More quantitative studies are required to better understand these features.

Combining the measurements of $R_{\rm GC}$--$\sigma_{\rm cc}$ and 
$R_{\rm GC}$--$\sigma_z$ for MCs near the tangent points, we can 
obtain the distribution of the total mass density in the Galactic plane 
(i.e., $R_{\rm GC}$--$\rho_0$ in Figure~\ref{density}). 
The error of $\rho_0$ can be calculated from the fitting error of $\sigma_z$
(Figure \ref{z0width}) and the assumed error of $\sigma_{\rm cc}$ (Figure \ref{sigv_12}).
We find that an exponential-disk model of $\rho_0(R)=\rho_{\rm GC}e^{-R/R_{\rm sl}}$
can roughly fit the distribution of the total mass density for the 
$R_{\rm GC}$=2.25--6.42~kpc region. 
Here, $\rho_{\rm GC}$=1.28~$\Msun$~pc$^{-3}$ is the mass density at the
Galactic center and $R_{\rm sl}=3.20$~kpc is the scale length of the disk.
The midplane mass density at the Sun is estimated to be
$\rho_0(R_0)$=0.10~$\Msun$~pc$^{-3}$, which agrees well with the Oort limit
summarized in \citet[][]{2015ApJ...814...13M}.
Because of the lower CO emission in the tangent points of $l\lsim22^{\circ}$ 
(i.e., $R_{\rm GC}\lsim$3~kpc for the 3 kpc ring structure in Figure~\ref{pv12}), 
we also fit the data in the $R_{\rm GC}$=3.05--6.42~kpc region. The resultant fitting of 
$R_{\rm GC}$--$\rho_0$ gives a similar result (see the blue line in Figure~\ref{density}).

\section{Summary}
\begin{figure}
\includegraphics[trim=0mm 0mm 0mm 0mm,scale=0.45,angle=0]{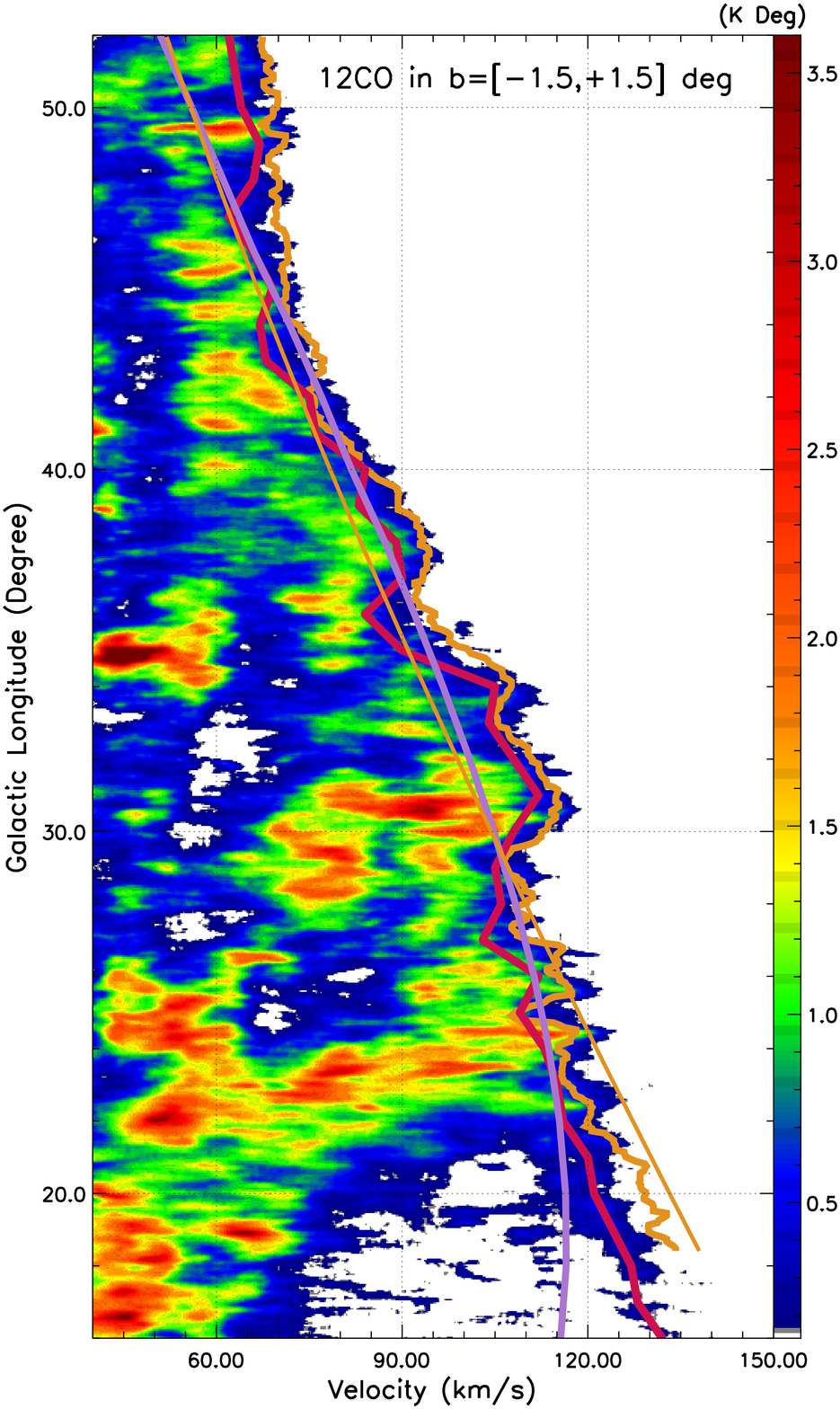}
\caption{
Longitude--velocity diagram of the \twCO\ emission for
the molecular gas along $l=16^{\circ}-52^{\circ}$. The region has a 
width of 3\fdg0, i.e., $b=[-$1\fdg5, 1\fdg5]. The red line indicates
the terminal velocity for the tangent-point gas ($V_{\rm tan}$) 
based on the MWISP CO data. The purple and golden lines
show the rotation curve from \citet[][]{Reid19} and 
the \HI\ result (the thick line for the measured values and the thin line 
for the linear fit) from \citet[][]{2016ApJ...831..124M}, respectively.
\label{pv12}}
\end{figure}

Based on the MWISP CO data, we have performed a study of the vertical
distribution of the molecular gas near the tangent points in the range of 
$l=16^{\circ}$--$52^{\circ}$ and $|b|\lsim$5\fdg1.
The main results are:

1. The terminal velocity of the molecular gas, which is comparable to
other observations and theoretical models, is well determined from
the new MWISP CO survey toward the inner Galaxy (Figure~\ref{pv12}). 

\begin{figure}
\includegraphics[trim=0mm 0mm 0mm 0mm,scale=0.43,angle=0]{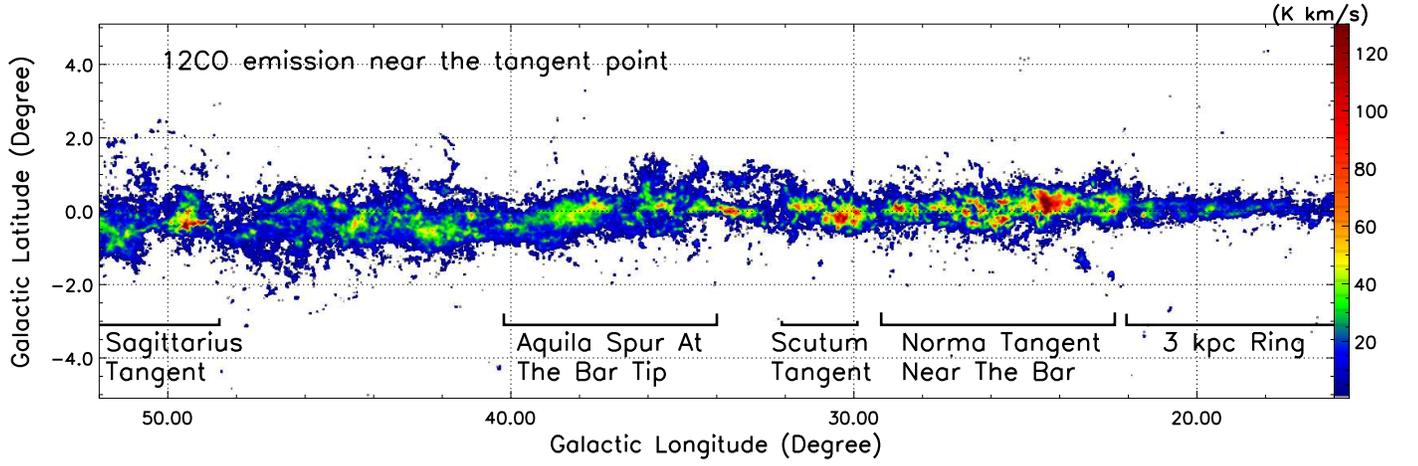}
\caption{
\twCO\ ($J$=1--0) intensity map toward $l=16^{\circ}-52^{\circ}$ 
in the velocity range of $V(l)\gsim V_{\rm tan}(l)-7$~km~s$^{-1}$. The tangent velocity,
i.e., $V_{\rm tan}(l)$, is well determined from the red line in Figure~\ref{pv12}.
Some large-scale structures seen in the longitude--velocity diagram 
(Figure~\ref{pv12}) are also labeled on the map of the integrated CO emission
near the tangent points 
\citep[e.g., see][]{2014ApJS..215....1V,2016ApJ...823...77R}.
\label{tan30cut}}
\end{figure}

\begin{figure}
\gridline{\fig{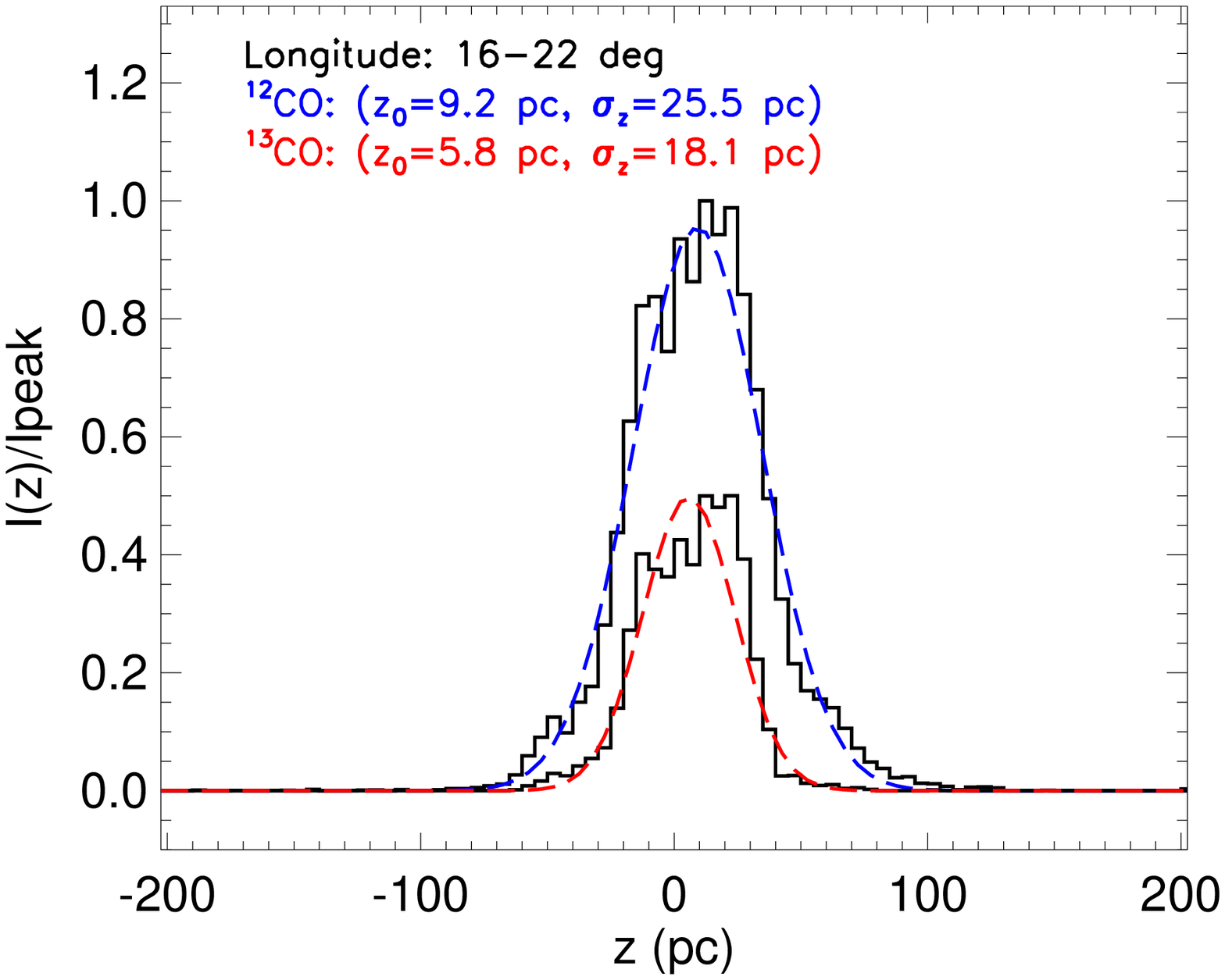}{0.45\textwidth}{(a)}
          \fig{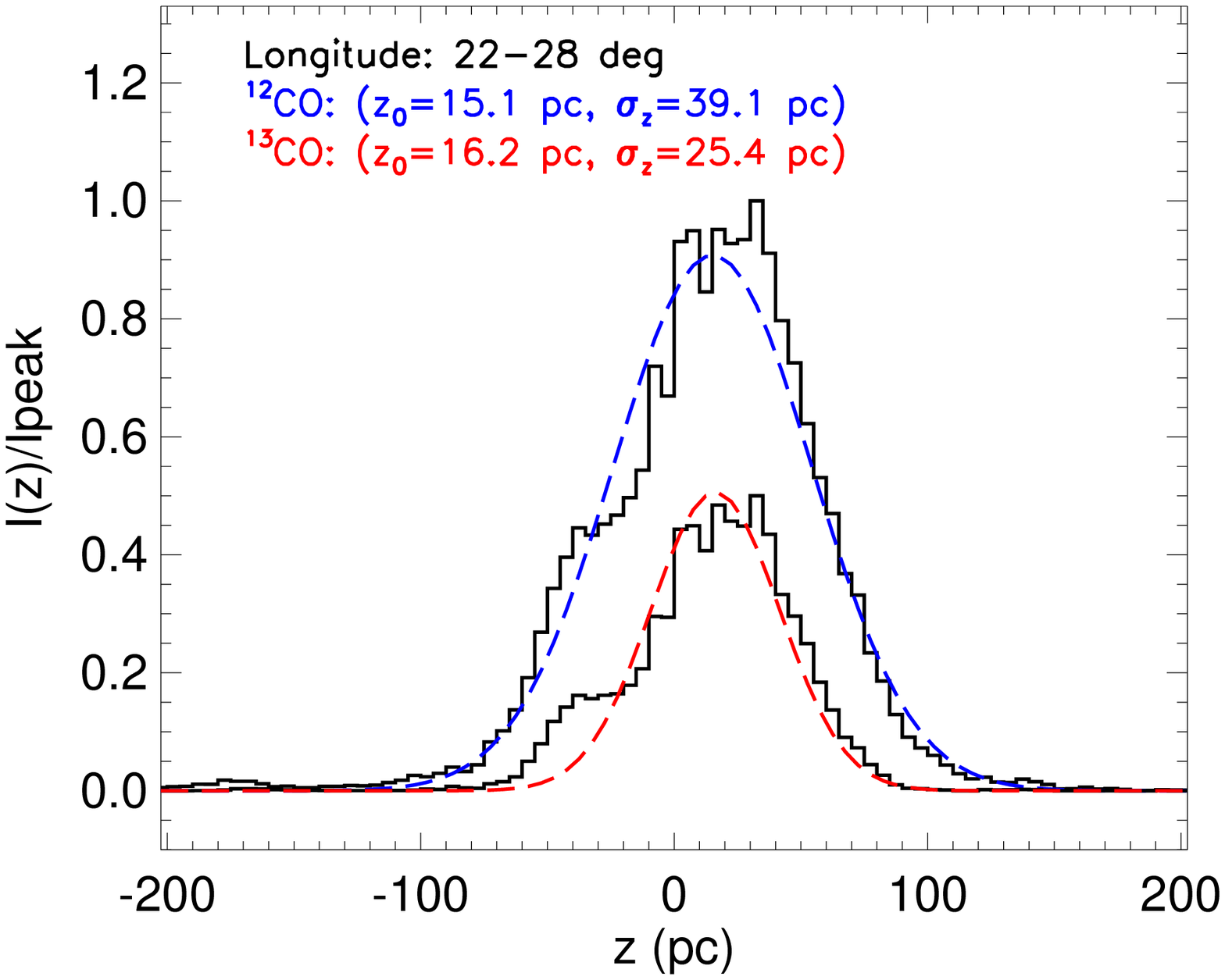}{0.45\textwidth}{(b)}
          }
\vspace{-5ex}
\gridline{\fig{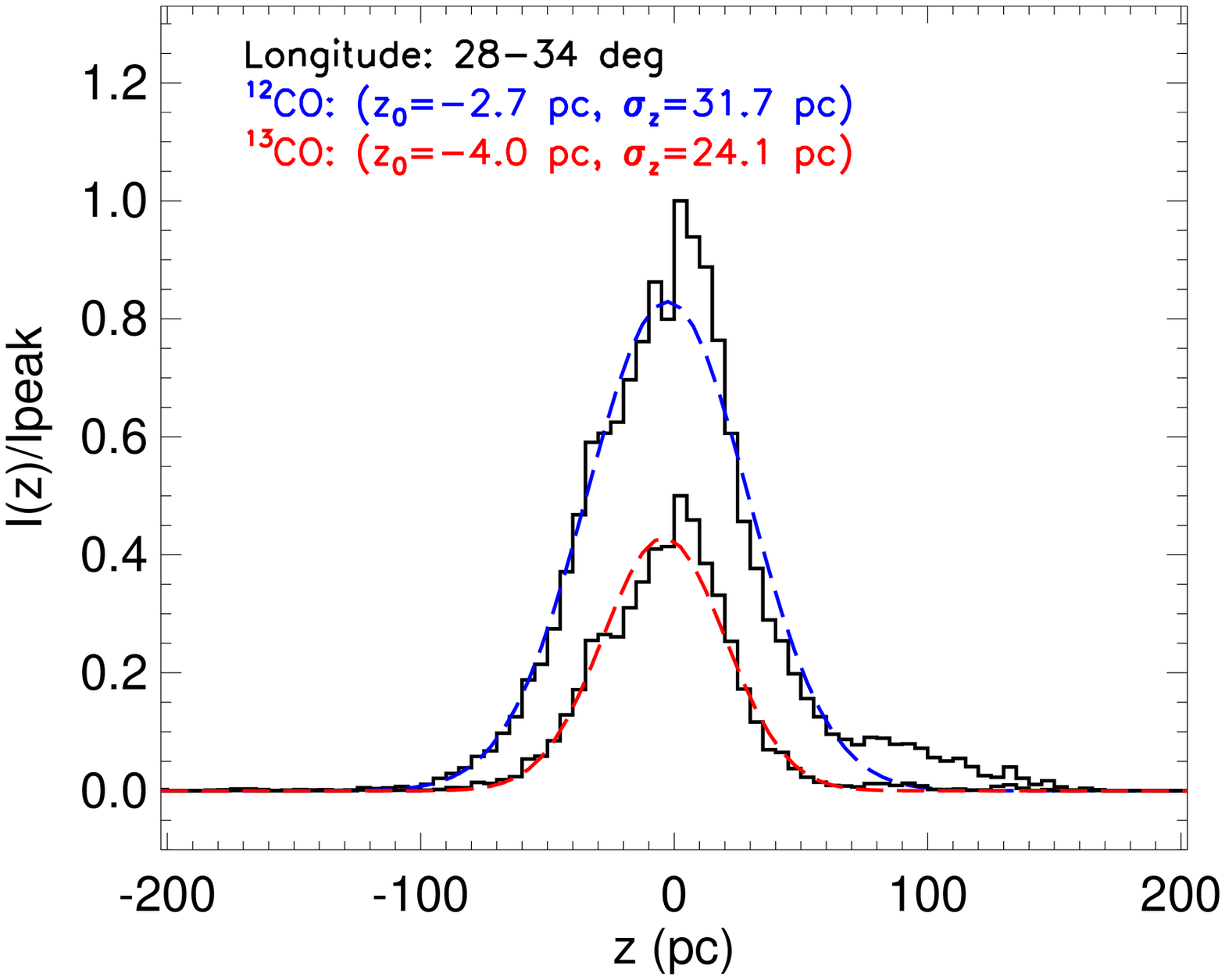}{0.45\textwidth}{(c)}
          \fig{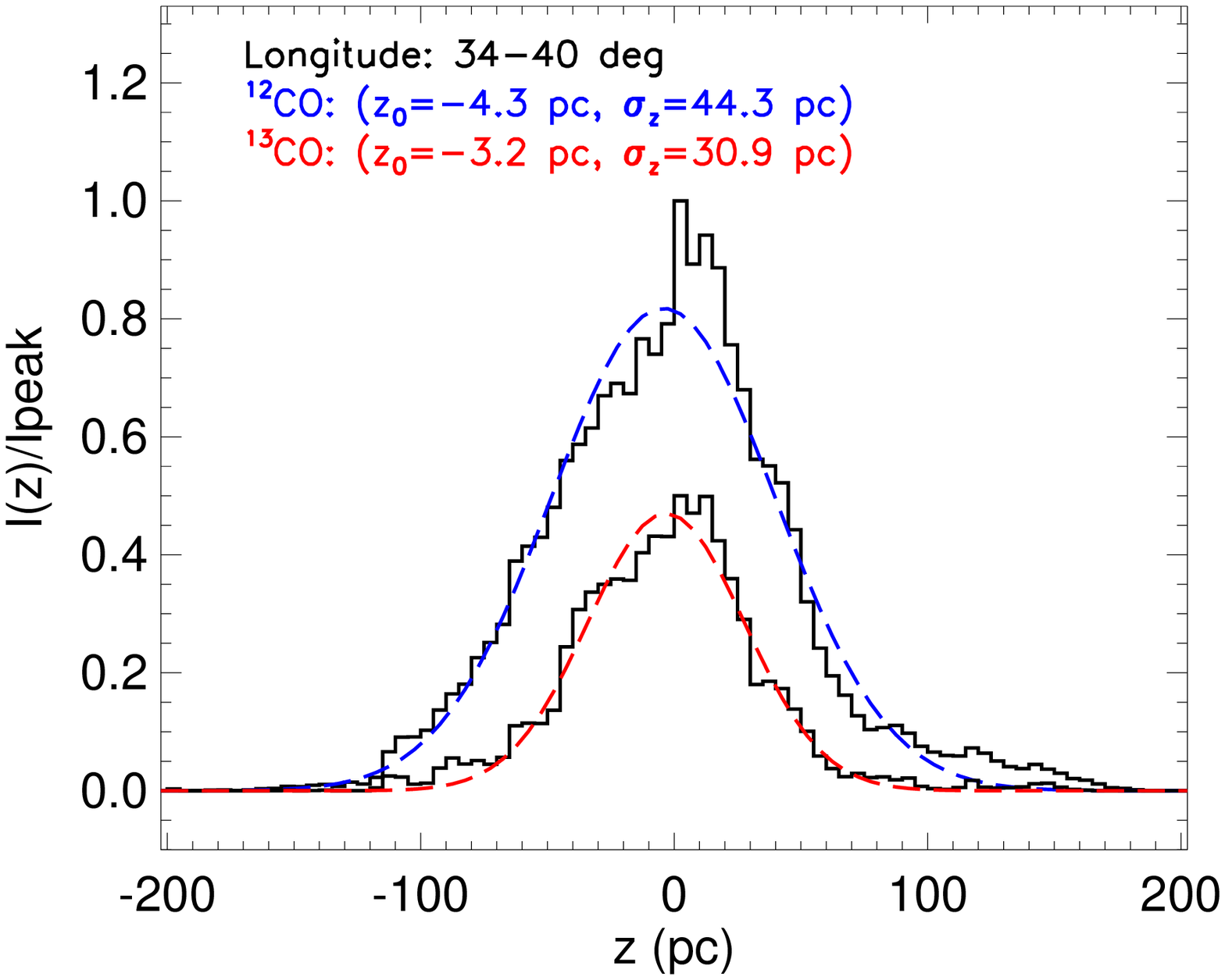}{0.45\textwidth}{(d)}
          }
\vspace{-5ex}
\gridline{\fig{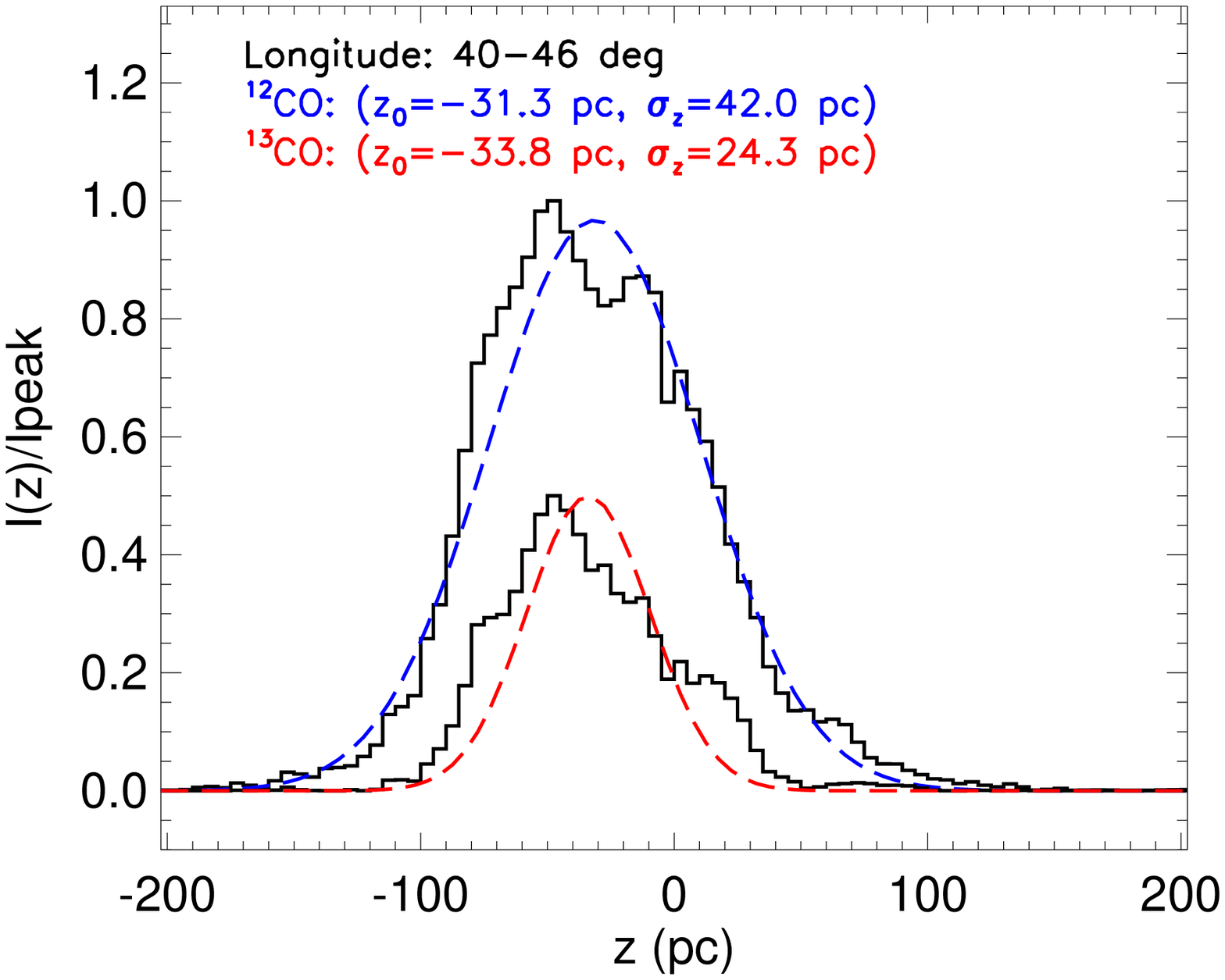}{0.45\textwidth}{(e)}
          \fig{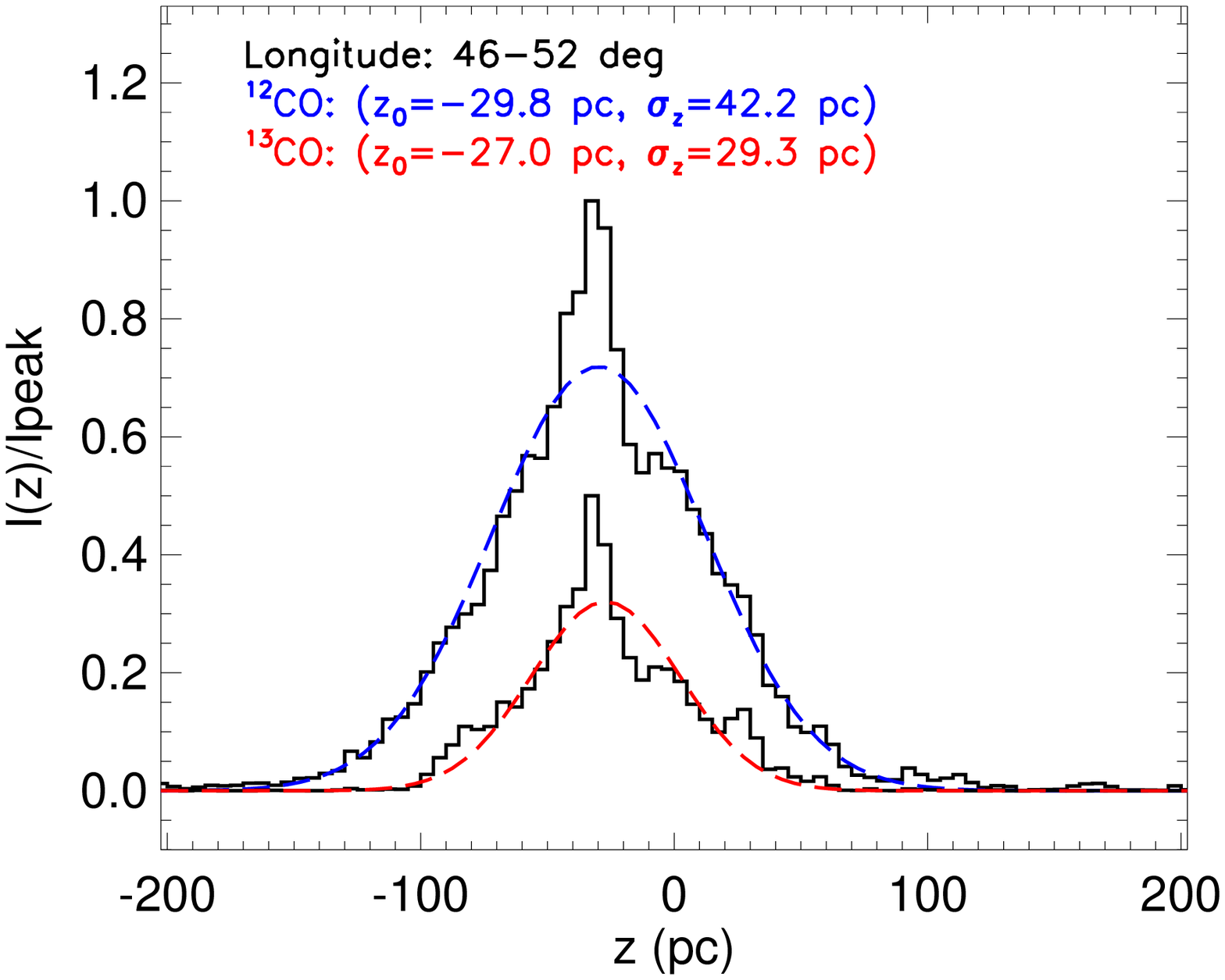}{0.45\textwidth}{(f)}
          }
\vspace{-2ex}
\caption{
(a)--(f) Vertical distribution of the total integrated 
\twCO\ (black solid line) and \thCO\ (black solid line, scale to 0.5)
intensity toward the longitude range of $16^{\circ}-22^{\circ}$, 
$22^{\circ}-28^{\circ}$, $28^{\circ}-34^{\circ}$, $34^{\circ}-40^{\circ}$, 
$40^{\circ}-46^{\circ}$, and $46^{\circ}-52^{\circ}$, respectively.
The bin in each panel is 5~pc, which corresponds to about 5 pixels (2\farcm5) 
at a distance of $\sim$7~kpc.
The blue dashed line indicates the Gaussian fit of the vertical
distribution for the Galactic thin disk based on the MWISP \twCO\ emission,
while the red dashed line is for the \thCO\ gas.
The fitted parameters of $z_0$ and $\sigma_z$ (i.e., FWHM=2.355$\times \sigma_z$) 
are also labeled on each panel for the one Gaussian component.
\label{narrow}}
\end{figure}

\begin{figure}
\gridline{\fig{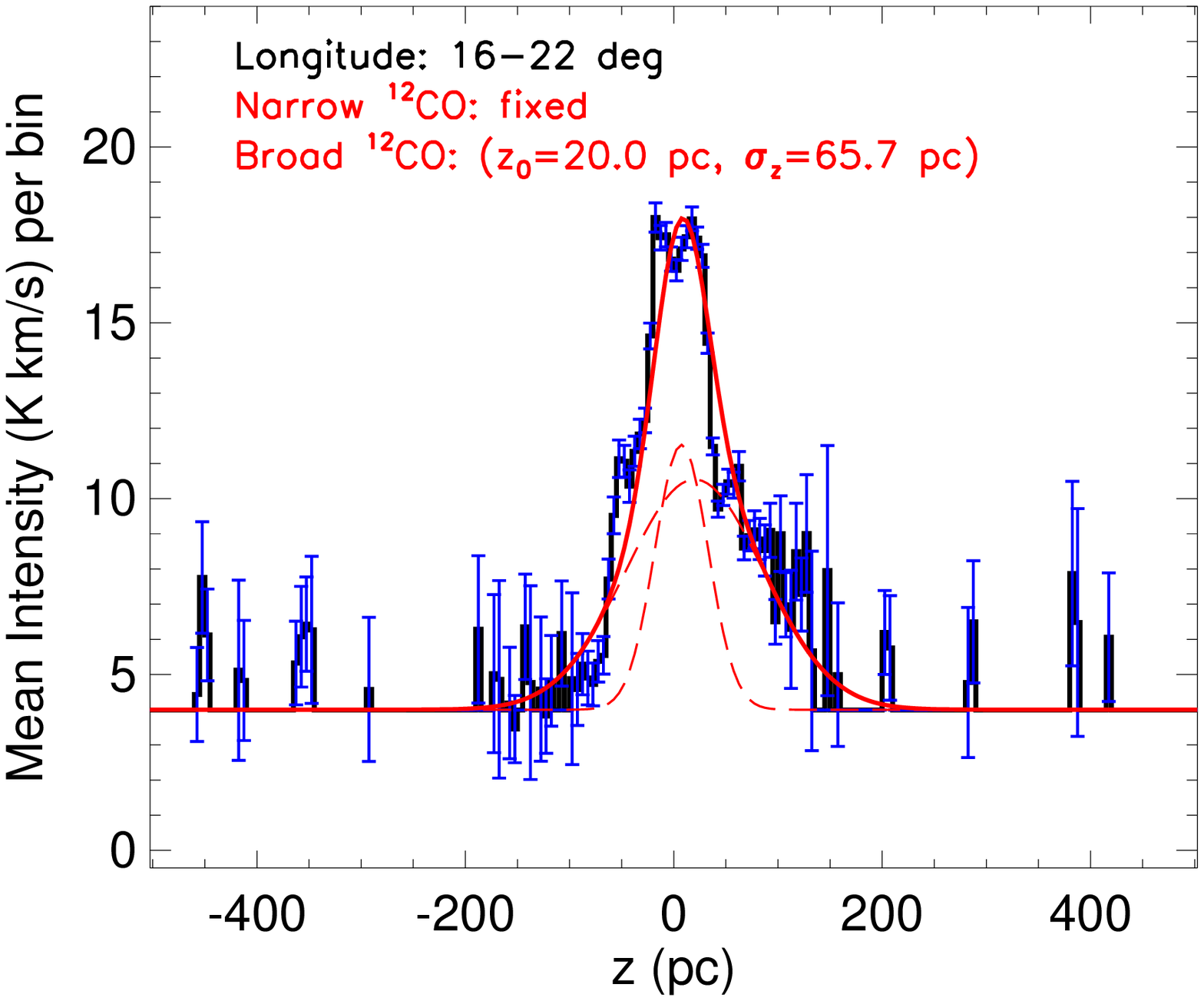}{0.45\textwidth}{(a)}
          \fig{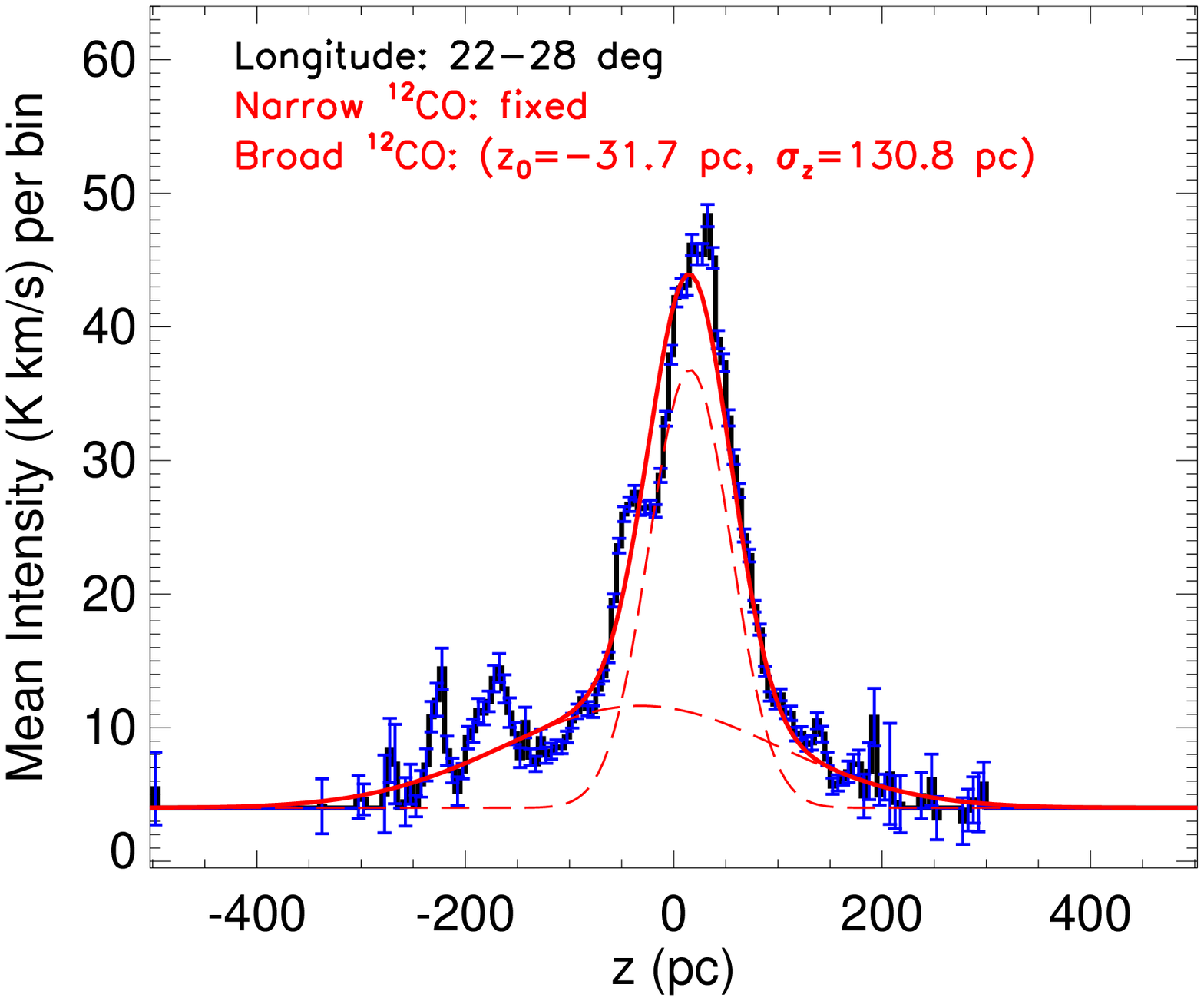}{0.45\textwidth}{(b)}
          }
\vspace{-5ex}
\gridline{\fig{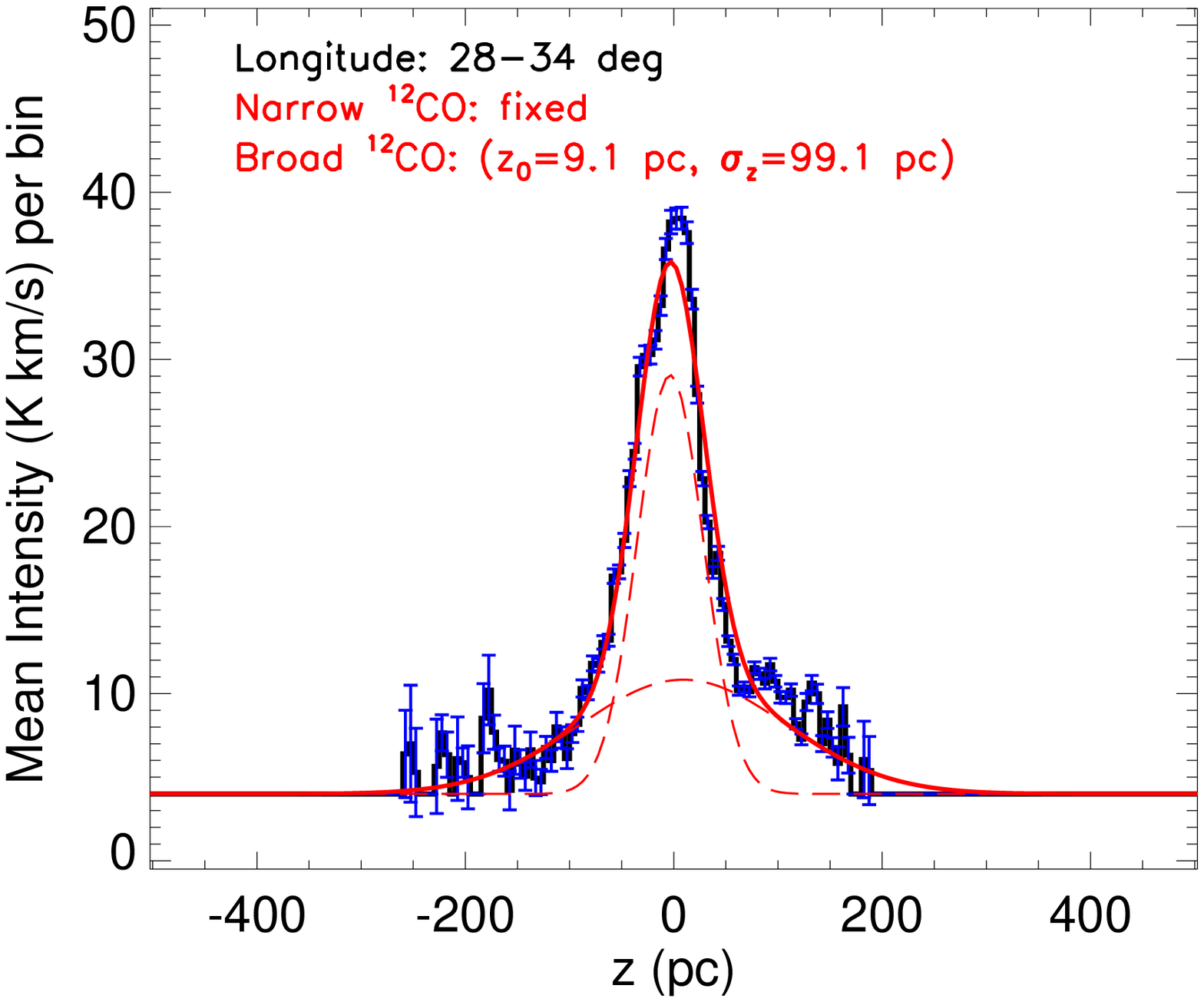}{0.45\textwidth}{(c)}
          \fig{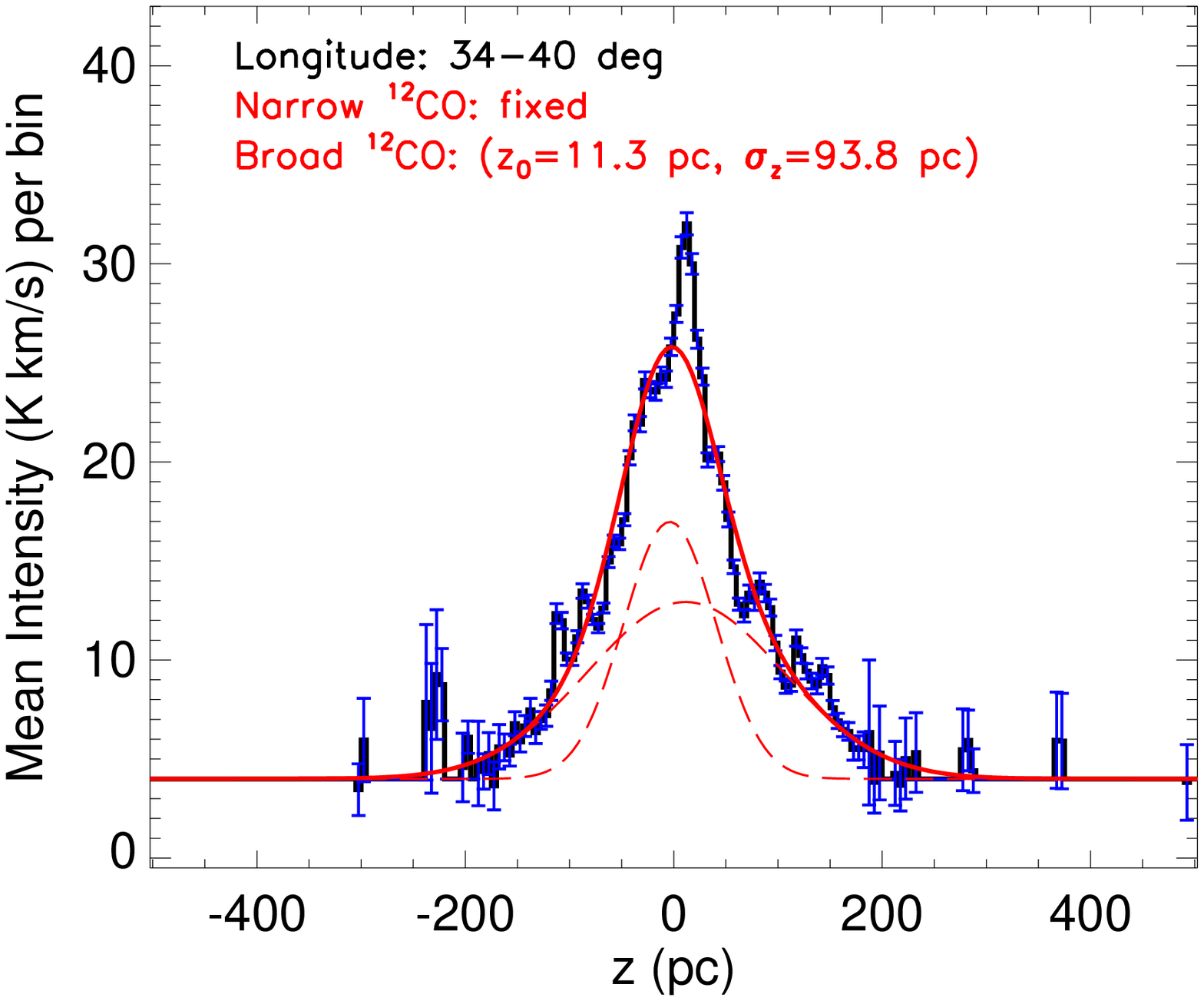}{0.45\textwidth}{(d)}
          }
\vspace{-5ex}
\gridline{\fig{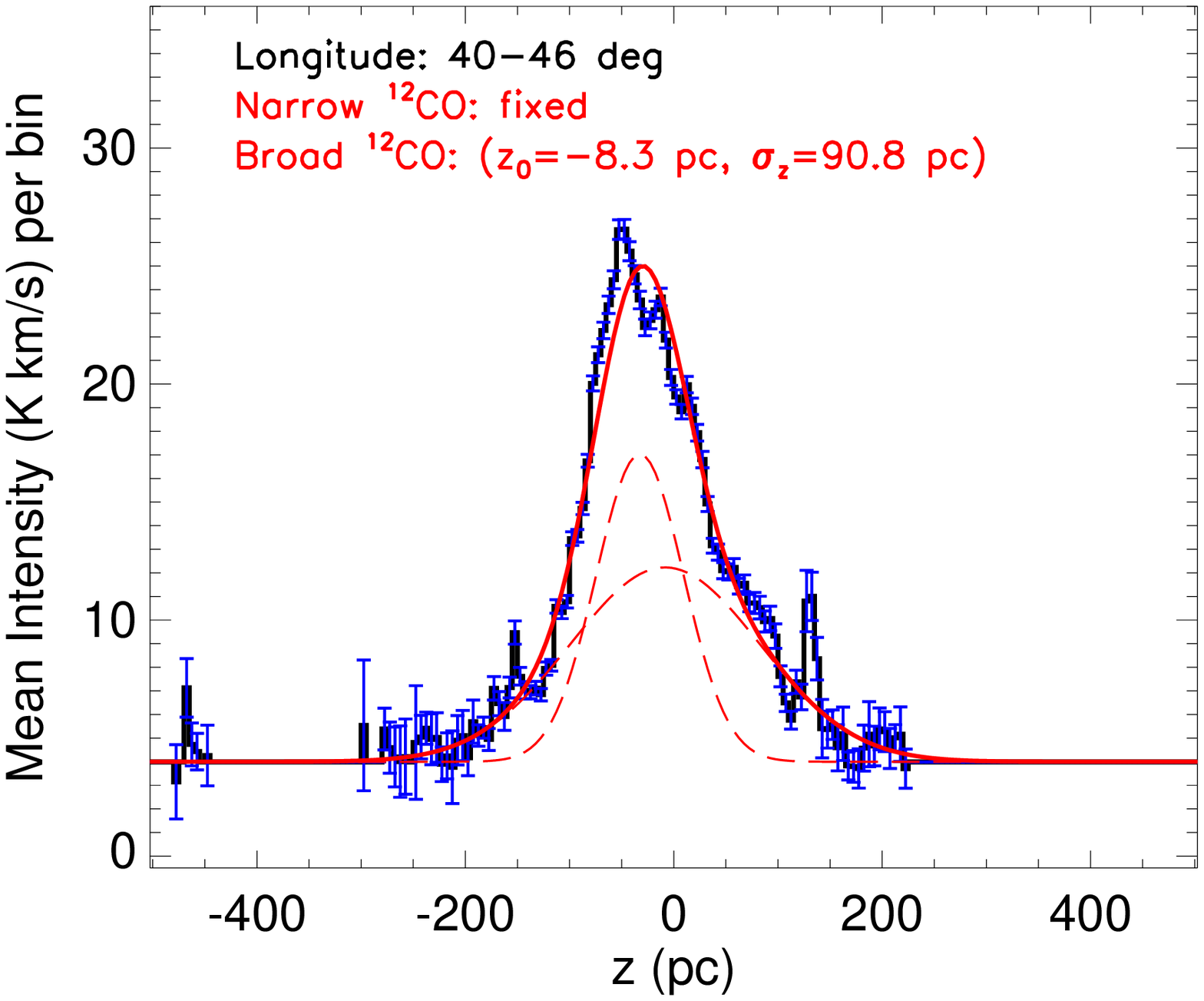}{0.45\textwidth}{(e)}
          \fig{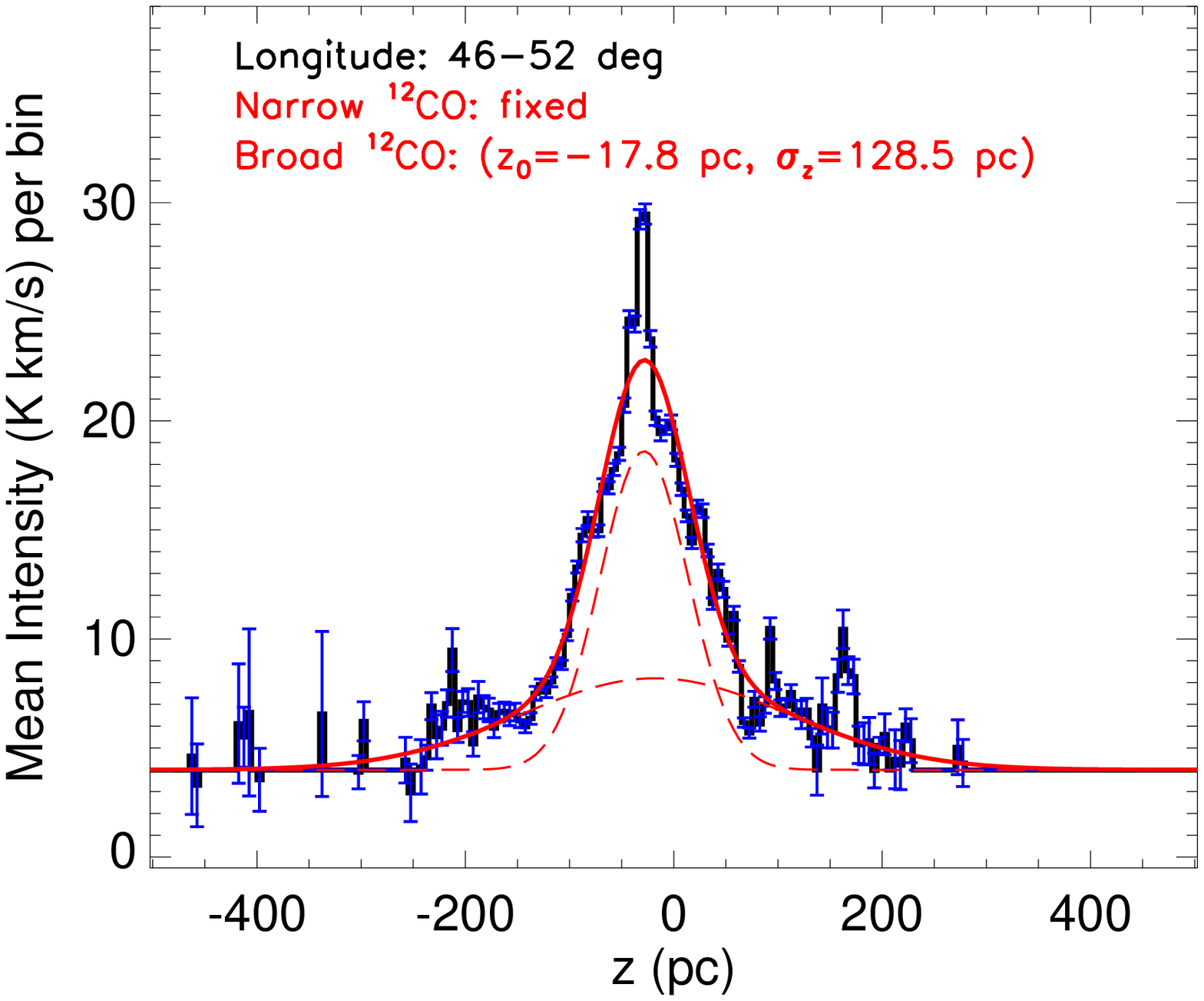}{0.45\textwidth}{(f)}
          }
\vspace{-2ex}
\caption{
(a)--(f) Vertical distribution of the mean intensity 
(i.e., the black solid line for $I_{\rm total}/\sqrt{N_{\rm pixel}}$, see the text)
of the \twCO\ emission toward the longitude range of $16^{\circ}-22^{\circ}$, 
$22^{\circ}-28^{\circ}$, $28^{\circ}-34^{\circ}$, $34^{\circ}-40^{\circ}$, 
$40^{\circ}-46^{\circ}$, and $46^{\circ}-52^{\circ}$, respectively. 
The bin in each panel is 5~pc. We assume $\sqrt{N_{\rm pixel}}$ errors for each bin.
The solid red line displays the best fit of two Gaussian components 
(i.e., the fixed narrow dashed line from Figure \ref{narrow} + the fitted 
broad dashed line) for the \twCO\ gas. 
The fitted parameters of $z_0$ and $\sigma_z$ (i.e., FWHM=2.355$\times \sigma_z$) 
are also labeled in each panel for the broad Gaussian component.
Note that the zero point of the fitting is roughly $4$~K~km~s$^{-1}$,
which is $\sim 3I_{\rm rms}$ of the integrated CO intensity near the tangent points.
\label{broad}}
\end{figure}

\begin{figure}
\gridline{\fig{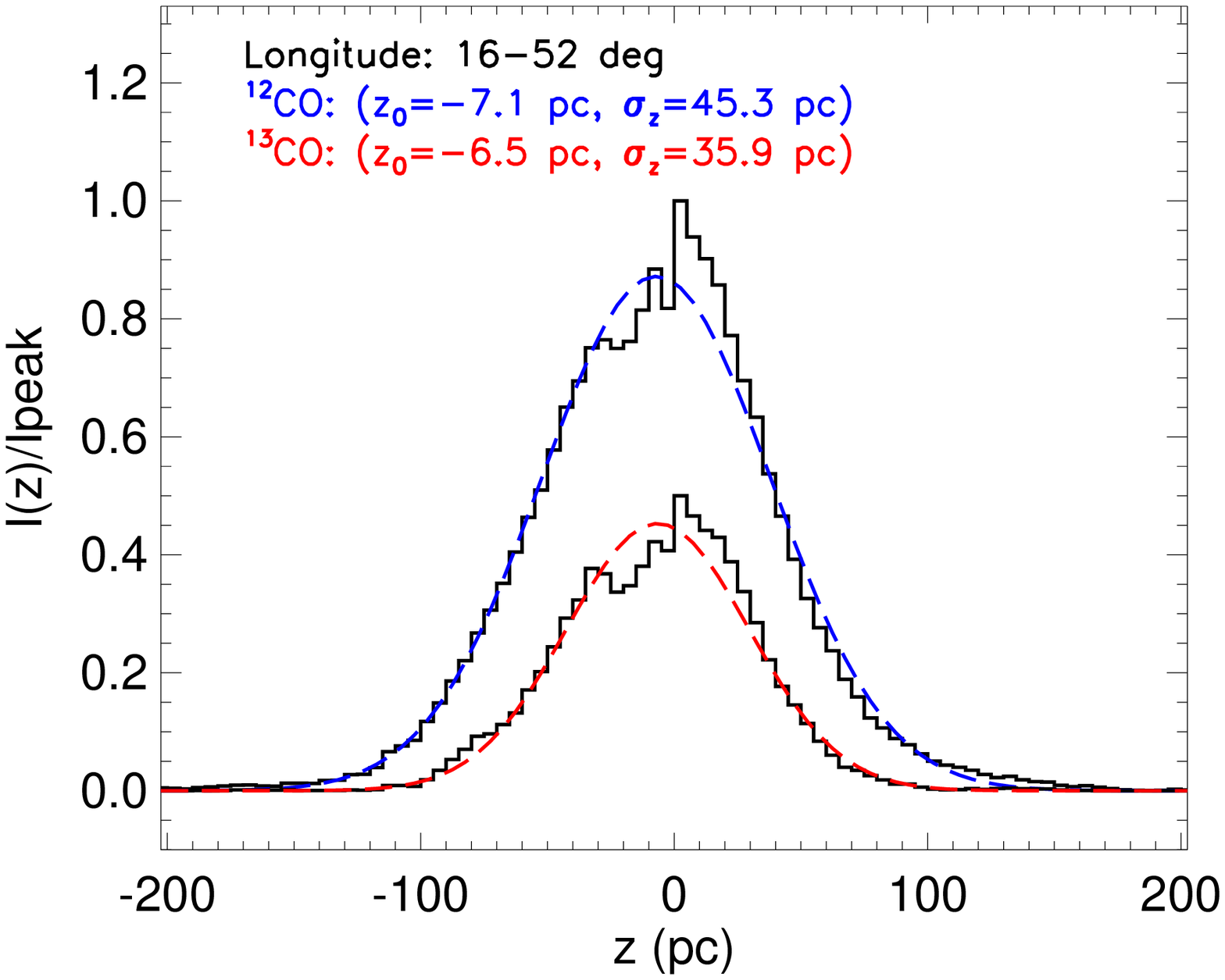}{0.45\textwidth}{(a)}
          \fig{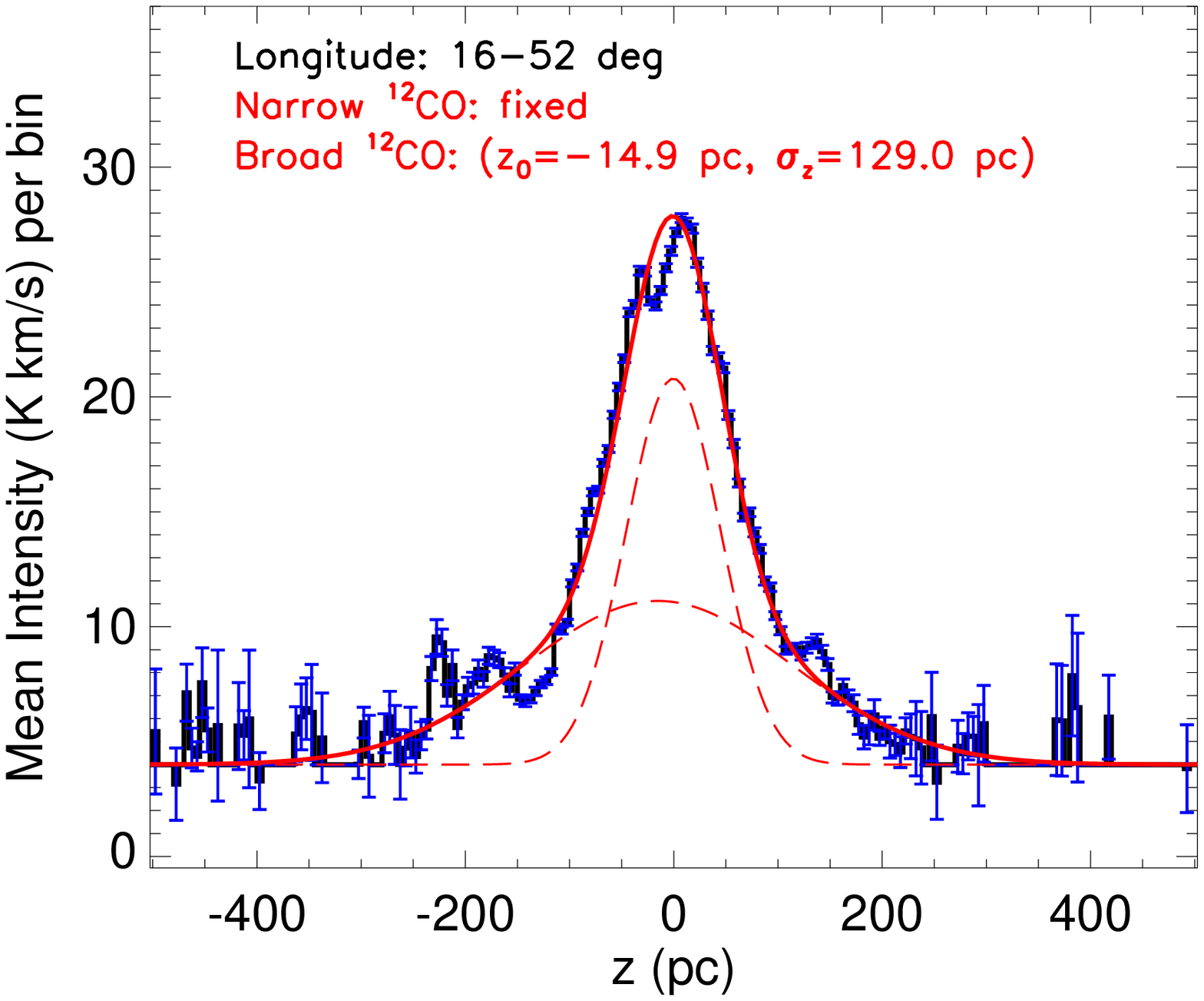}{0.45\textwidth}{(b)}
          }
\vspace{-2ex}
\gridline{\fig{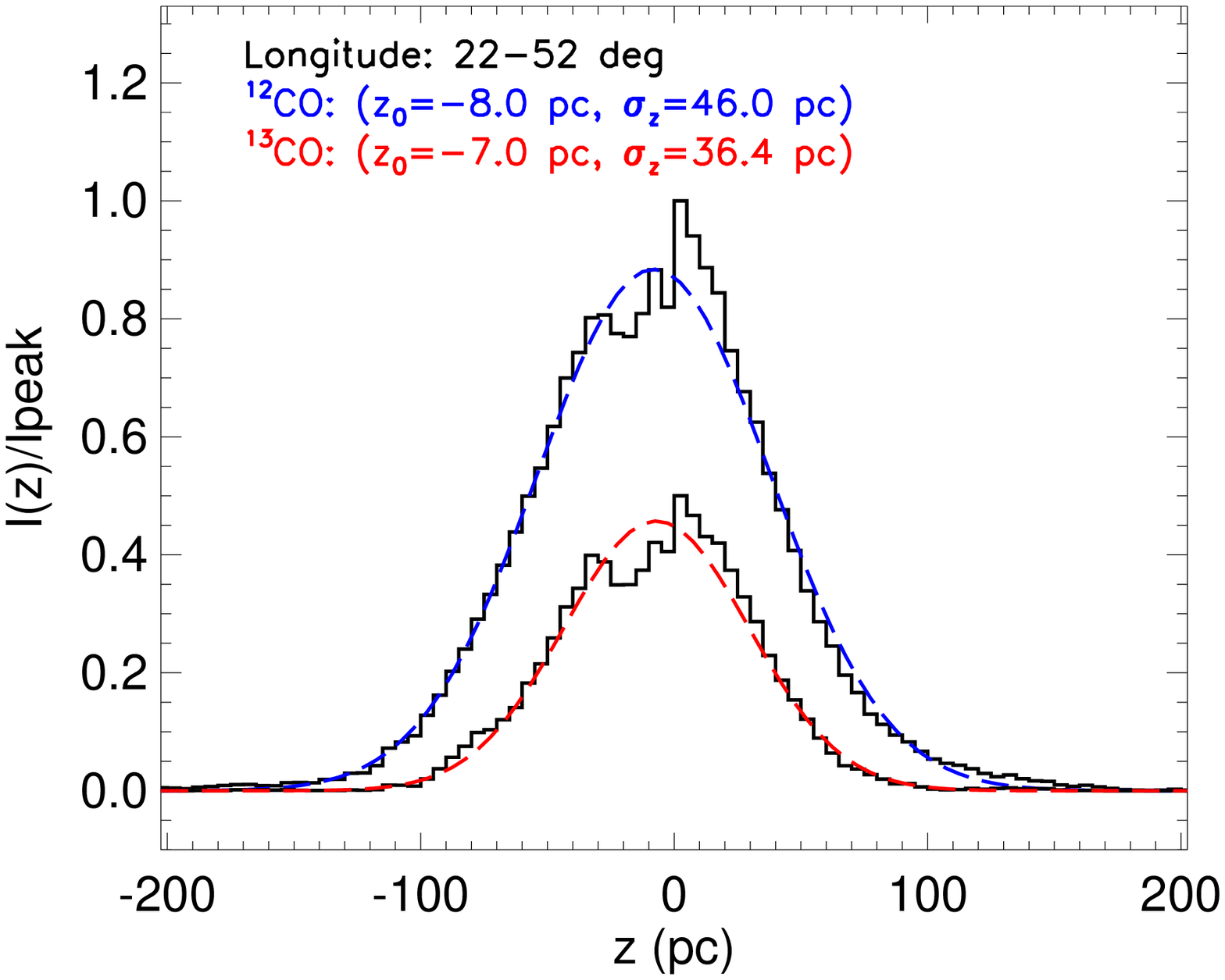}{0.45\textwidth}{(c)}
          \fig{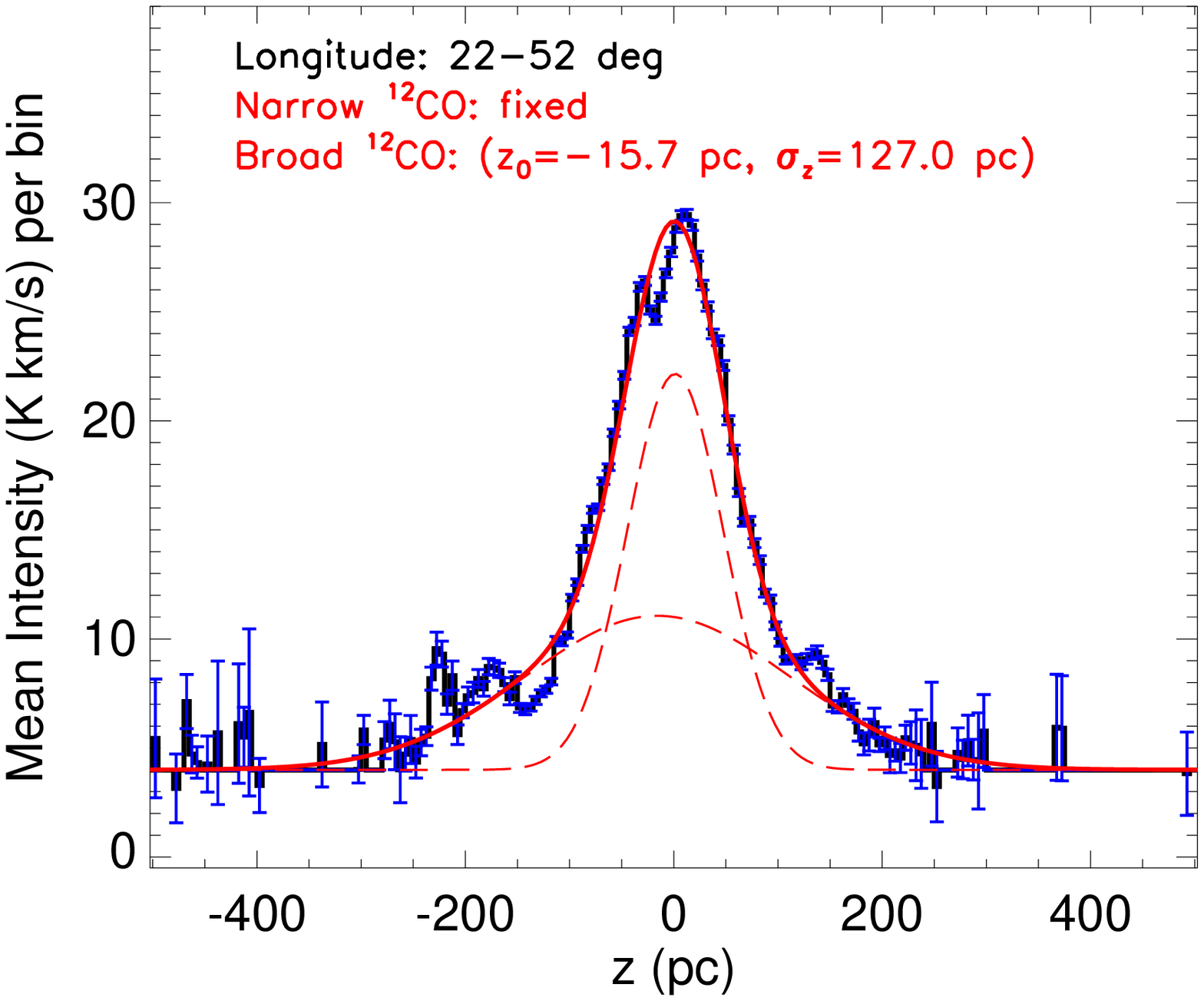}{0.45\textwidth}{(d)}
          }
\vspace{-2ex}
\caption{
Left panels: same as Figure \ref{narrow} but for the longitude range 
of $16^{\circ}-52^{\circ}$ and $22^{\circ}-52^{\circ}$ based on 
the total integrated CO emission. The fitted parameters of the \twCO\
emission for the thin disk are used to fix the narrow Gaussian component 
in the right panels.
Right panels: same as Figure \ref{broad} but for the longitude range
of $16^{\circ}-52^{\circ}$ and $22^{\circ}-52^{\circ}$ based on 
the mean intensity of the \twCO\ emission.
The fitted parameters of $z_0$ and $\sigma_z$ (i.e., FWHM=2.355$\times \sigma_z$)
are labeled in each panel for the broad Gaussian component (i.e., the thick CO disk).
\label{disk1652}}
\end{figure}

2. We use the CO emission near the terminal velocity to trace the molecular gas 
distribution at the tangent points, which can avoid the distance ambiguity within 
the solar circle. The high-quality CO data reveal two main molecular gas features 
near the tangent points, including large-scale bright CO emission in the Galactic midplane 
and discrete MCs with very weak emission in relatively 
high-$b$ regions (see Figure~\ref{tan30cut}).

\begin{figure}
\gridline{\fig{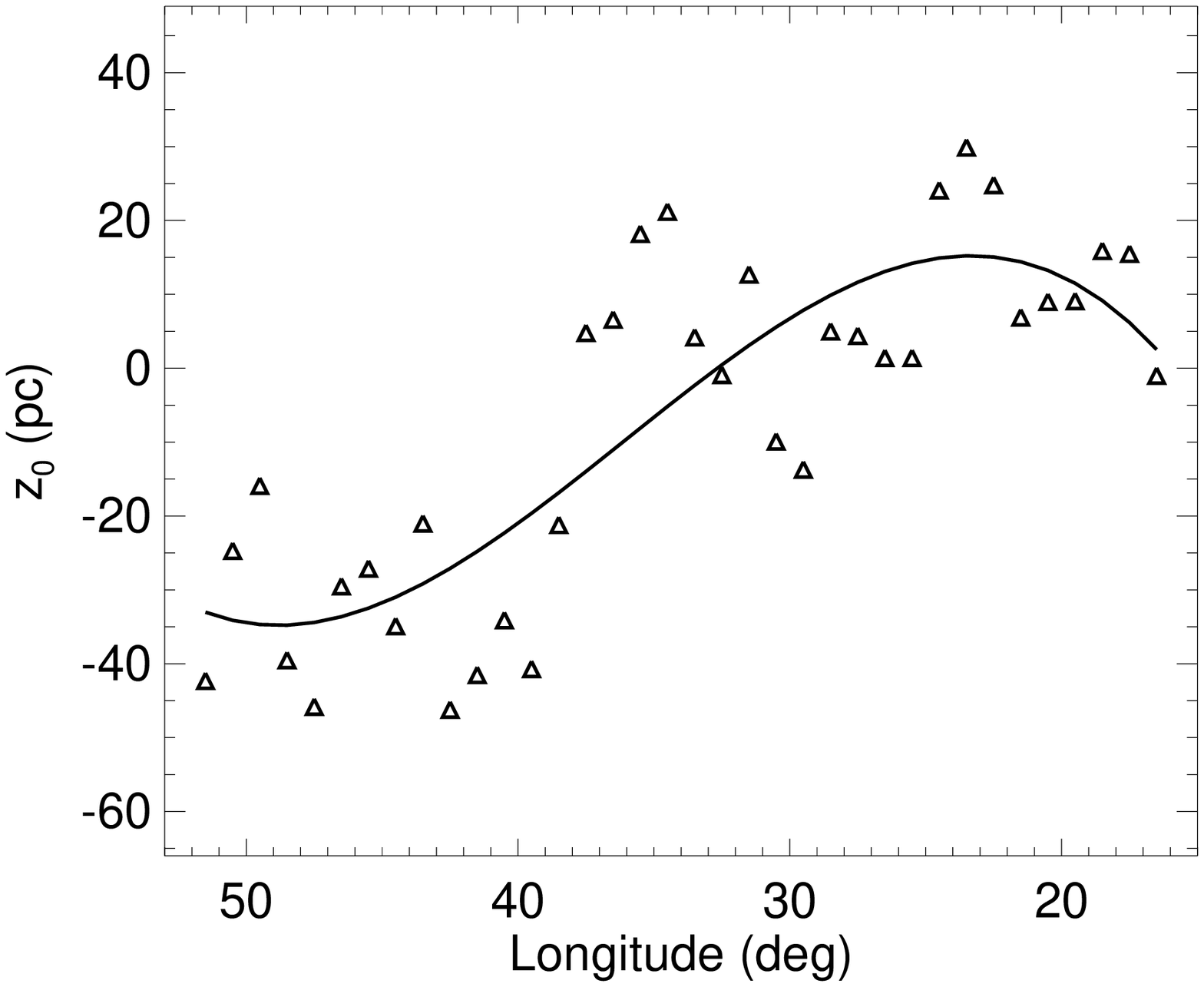}{0.5\textwidth}{(a)}
          \fig{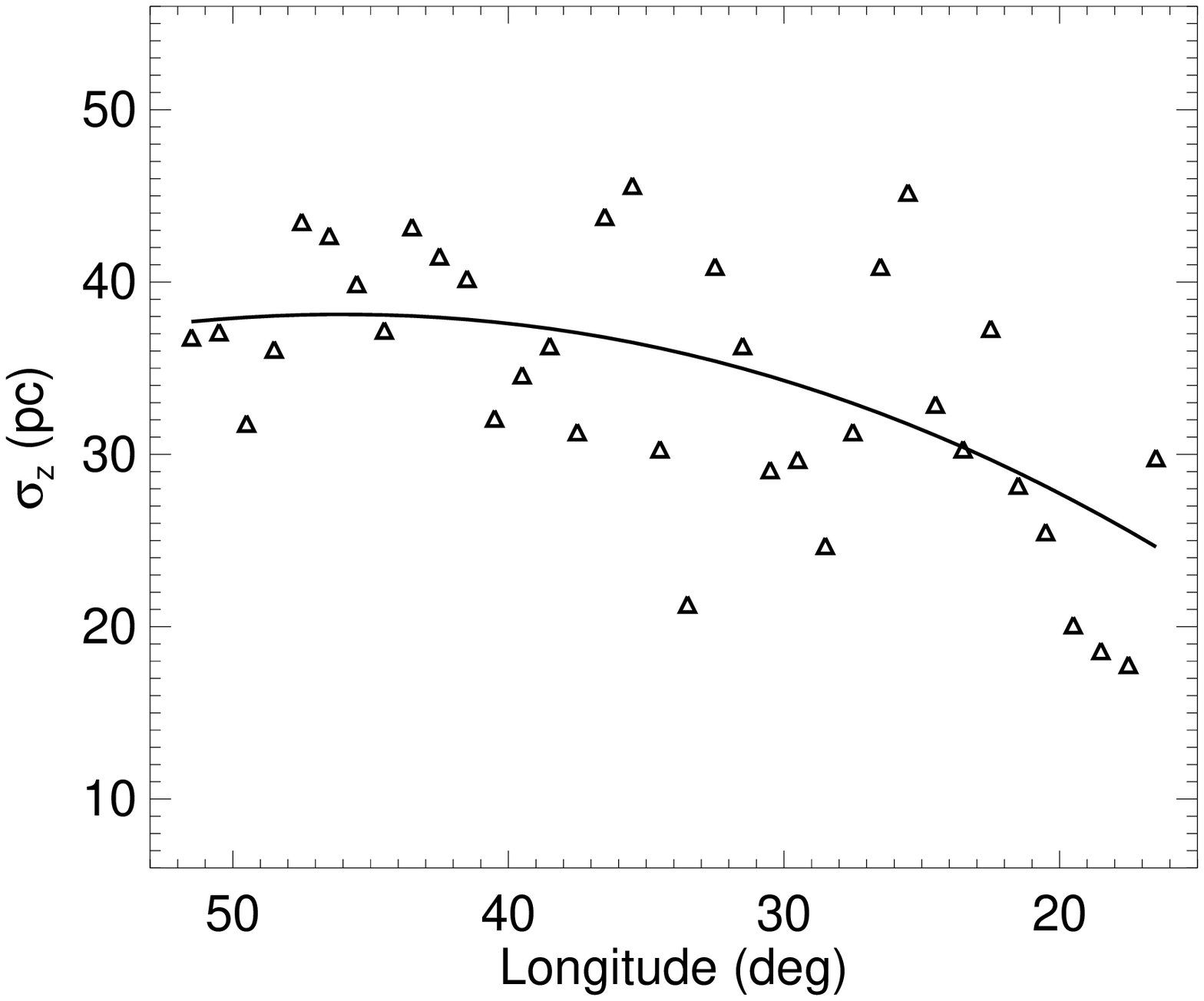}{0.5\textwidth}{(b)}
          }
\caption{
The position of the centroid of the molecular gas near the tangent points 
($z_0$ in the left panel) and the scale height of the thin gas disk 
($\sigma_{z}$ in the right panel) based on the \twCO\ emission
at different Galactic longitudes of 
$l=16^{\circ}-52^{\circ}$ with a separation of $1^{\circ}$.
The line indicates the least-square fit from a third-order polynomial
by using the 36 points.
\label{z0width}}
\end{figure}

\begin{figure}
\includegraphics[trim=0mm 0mm 0mm 0mm,scale=0.6,angle=0]{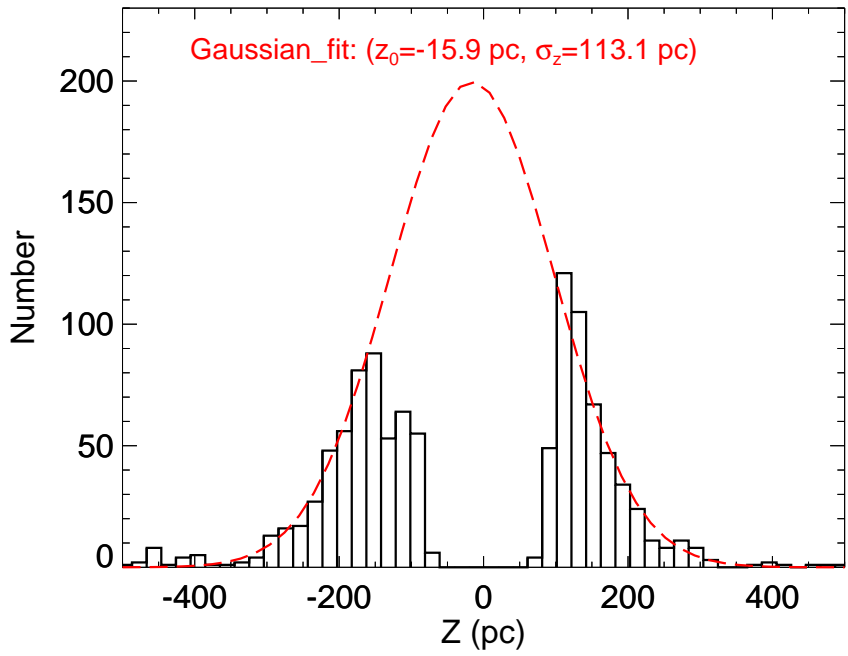}
\caption{
Vertical distribution of the identified MCs far from the Galactic plane.
The red dashed line shows the Gaussian fit for the broad component
of the Galactic thick disk.
\label{sta_zmass}}
\end{figure}

3. Based on the integrated CO intensity near the terminal velocity, we find that
the model of two Gaussian components can be used to describe the vertical distribution
of the molecular gas near the tangent points.
The narrow one with a FWHM of $\sim$85~pc is the well-known thin molecular gas disk
in the inner Galaxy, while another broad component with a FWHM of $\sim$260--300~pc
probably represents the thick CO disk. 
For the thin CO disk, Figure~\ref{z0width} displays
the systematic variation of the offset ($z_0$) and the scale height ($\sigma_z$)
of the molecular gas with respect to the Galactic longitude, which is consistent with 
previous studies \citep[e.g.,][]{Malhotra94}.
For the thick CO disk, its thickness is
about three times as wide as that of the well-known thin molecular gas layer,
and the thickness of the thick molecular gas layer is very comparable in 
width to the central atomic gas layer \citep[e.g., 250-270~pc in][]
{Dickey90,Lockman91}. 

4. For the thick CO disk, a total of 1055 high-$z$ MCs were identified at the 
tangent points in the first quadrant region of the MWISP survey. 
These MCs have a median radius of 2.5~pc, a median velocity dispersion of
0.8~$\km\ps$, and a median peak temperature of 2.1~K. The mass surface brightness
of the MCs is very low, i.e., a median value of 6.8~$\Msun {\rm pc}^{-2}$
assuming a constant CO-to-H$_2$ conversion factor of
$2\E{20}$~cm$^{-2}$(K~km~s$^{-1})^{-1}$.
The cloud-cloud velocity dispersion of the high-$z$ MCs is
4.4--5.6~$\km\ps$, which is similar to the value of 4.9$\pm1.3\ \km\ps$
for all MCs at the tangent points.
The high virial parameter indicates that the high-$z$ MCs
are probably short-lived objects. That is, 
the MCs of the new disk population are either dispersing or
being assembled by some external dynamical processes.
Alternatively, some of the high-$z$ MCs are probably newborn clouds
in an environment of the high ram pressure.
Our findings show that the thick CO disk is composed of many
discrete MCs with small size and low mass 
(i.e., $\frac{dN}{dM}\propto M^{-1.74}$ in the MC's mass 
range of $\sim120-8000\ \Msun$).

5. Nearly 90\% of these MCs are $\gsim$100~pc from the plane. However, 
only 3\% of MCs are in the $|z|\gsim360$~pc region, suggesting that 
the high-$z$ molecular gas is the disk population. 
Including the emission of the CO layer through the whole Galactic
plane, the total molecular gas mass of the thick CO disk is estimated to be
$8.5\times10^7\ \Msun$ in the range of $R_{\rm GC}=$2--8.15~kpc,
which is about 10\% of the total molecular mass in the same region of 
the inner Galaxy.
The surface density and midplane mass density of the thick molecular 
gas disk are $\sim0.4\ \Msun {\rm pc}^{-2}$ and $\sim0.0014\ \Msun {\rm pc}^{-3}$
(or $\sim0.02\ {\rm H_2}{\rm cm}^{-3}$), respectively.

6. The cloud-cloud velocity dispersion of all tangent MCs seems to be smaller 
at larger $R_{\rm GC}$, leading to a slope of $\sim-0.4\ \km\ps$kpc$^{-1}$ 
in the region of $R_{\rm GC}$=2.2--6.4~kpc. This value is about one-half 
of the \HI\ gas when the joint gravitational potential is considered
\citep[][]{2002A&A...394...89N}.
The higher star formation rate at smaller $R_{\rm GC}$ is probably
responsible for this feature.

7. Assuming the vertical equilibrium between the turbulent pressure of the 
molecular gas and the total gravitational force in the disk, we can use an 
exponential-disk model of $\rho_0(R)=\rho_{\rm GC}e^{-R/R_{\rm sl}}$ to fit 
the distribution of the mass density in the Galactic midplane. The best-fit
parameters are $R_{\rm sl}$=3.2~kpc and $\rho_{\rm GC}$=1.28~$\Msun {\rm pc}^{-3}$.
The midplane mass density in the solar neighborhood is estimated to be 
0.10~$\Msun {\rm pc}^{-3}$ at $R_0=$8.15~kpc, which agrees with 
the local mass density of 0.097$\pm0.013\ \Msun {\rm pc}^{-3}$ 
suggested by \citet[][]{2015ApJ...814...13M}.

\acknowledgments
We are grateful to all the members of the MWISP working group, particularly 
the staff members at the PMO-13.7m telescope, for their long-term support. 
We also thank the anonymous referee for providing many helpful comments and 
suggestions that largely improved the paper. MWISP was sponsored by National 
Key R\&D Program of China with grant no. 2017YFA0402700 and CAS Key Research 
Program of Frontier Sciences with grant no. QYZDJ-SSW-SLH047. J.Y. and X.C. are
supported by National Natural Science Foundation of China through grant 12041305. 
Y.S. is supported by the Youth Innovation Promotion Association, CAS (2018355).
This work is supported by NSFC 11629302, 11773077, and 11803091.

\facility{PMO 13.7m}
\software{GILDAS/CLASS \citep{2005sf2a.conf..721P}} 

\bibliographystyle{aasjournal}
\bibliography{references}

\clearpage

\clearpage

\clearpage

\clearpage

\clearpage

\clearpage

\clearpage

\clearpage

\begin{figure}
\gridline{\fig{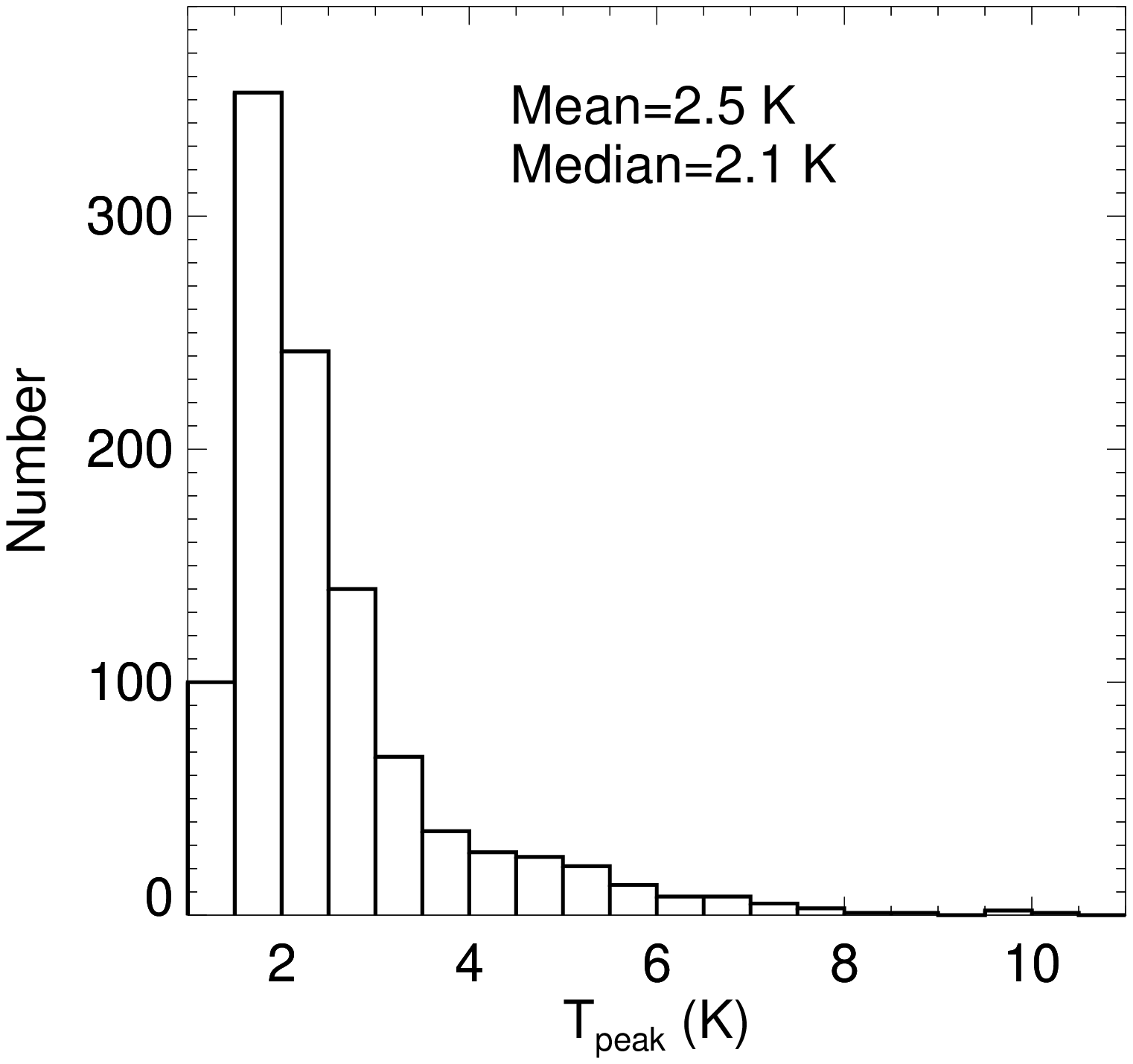}{0.35\textwidth}{(a)}
          \fig{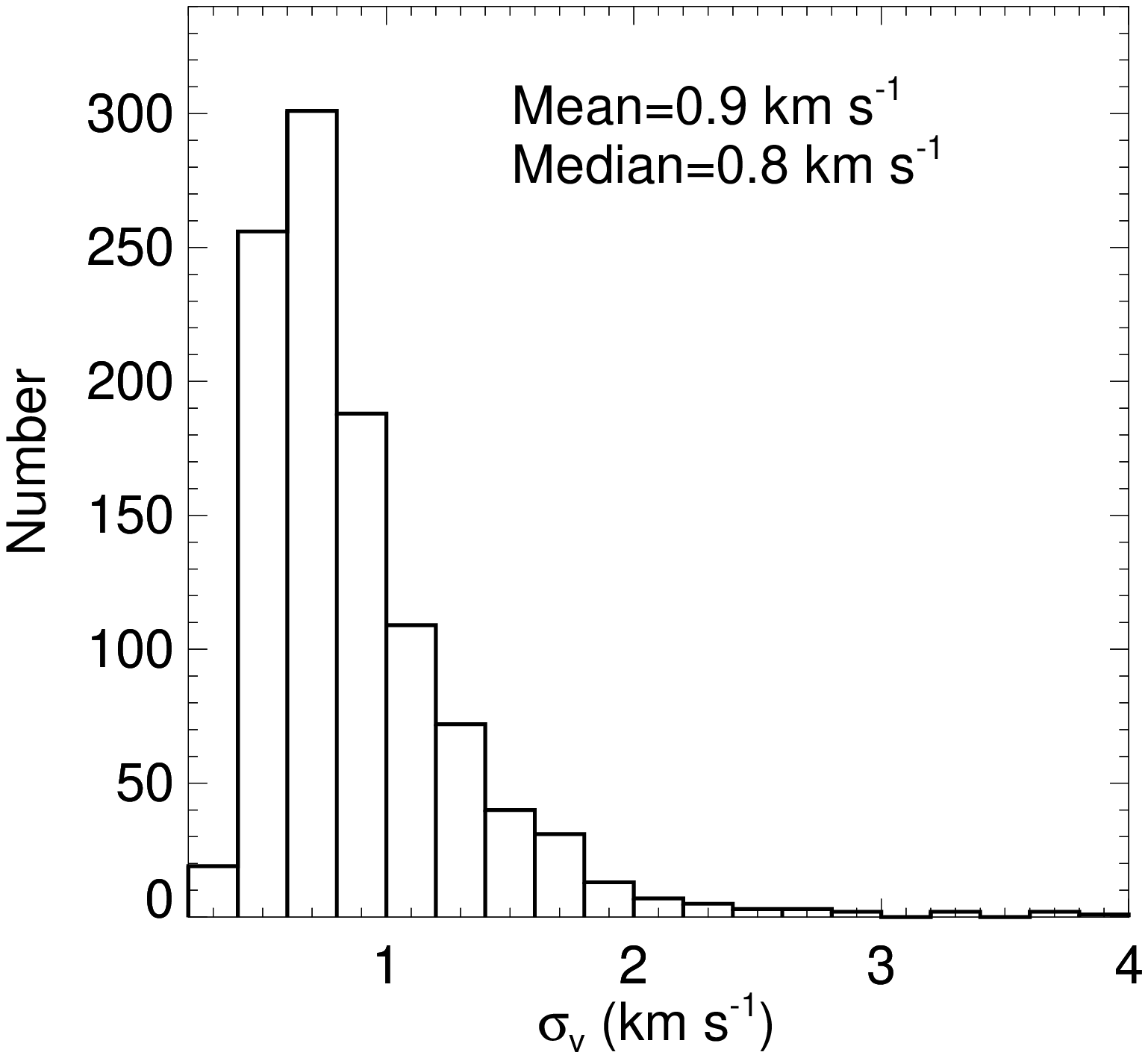}{0.35\textwidth}{(b)}
          }
\vspace{-5ex}
\gridline{\fig{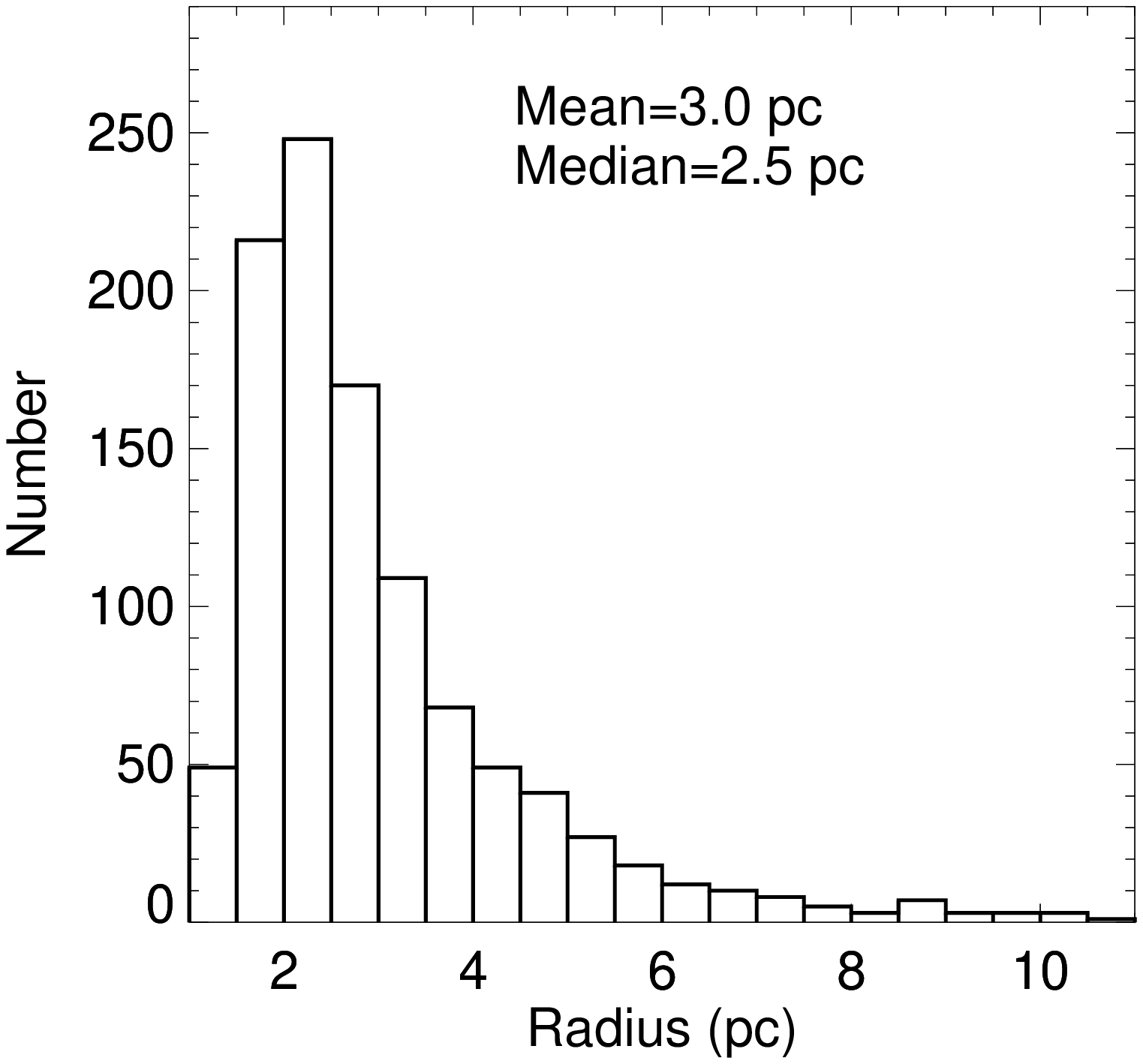}{0.35\textwidth}{(c)}
	  \fig{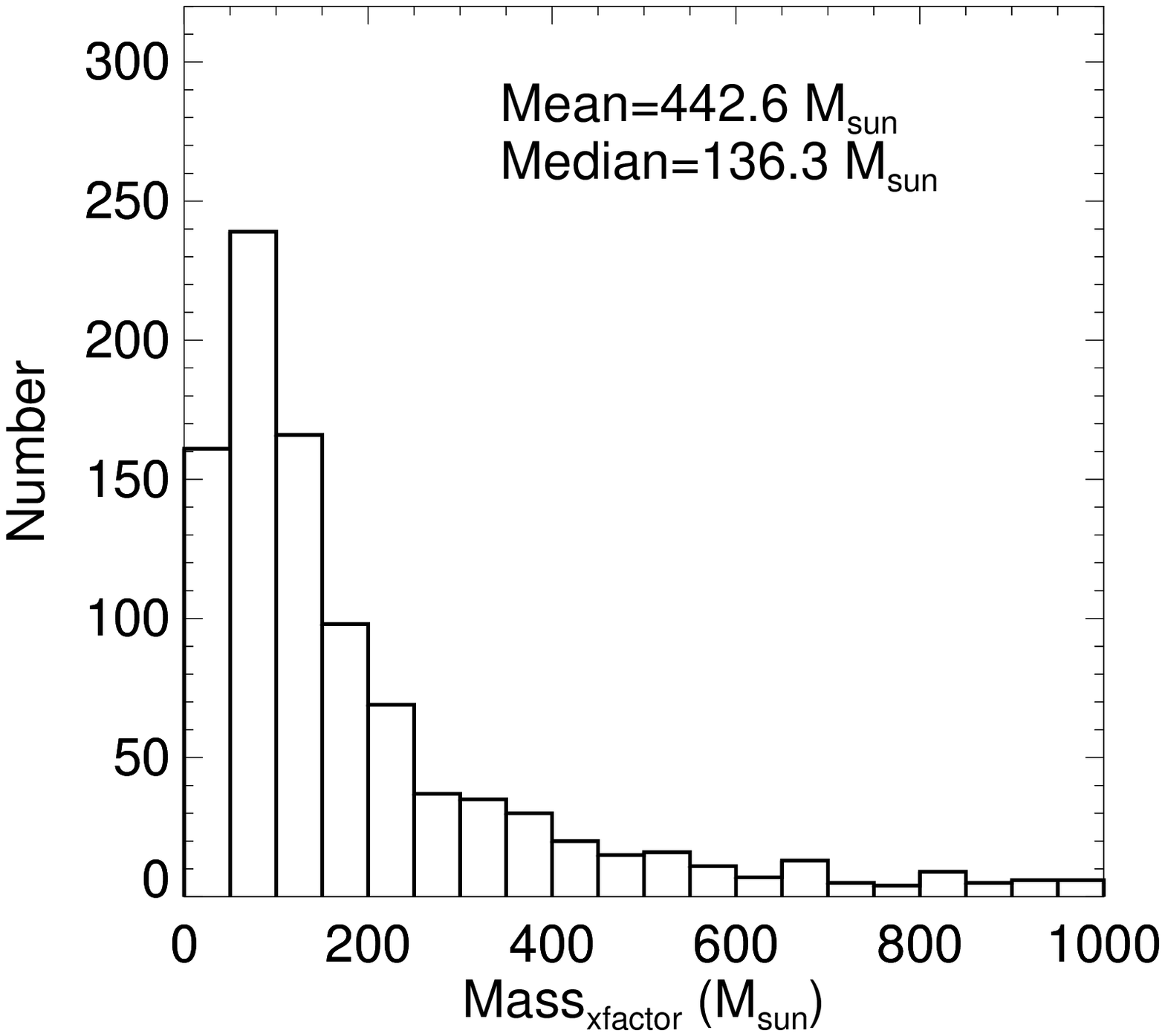}{0.35\textwidth}{(d)}
          }
\vspace{-5ex}
\gridline{\fig{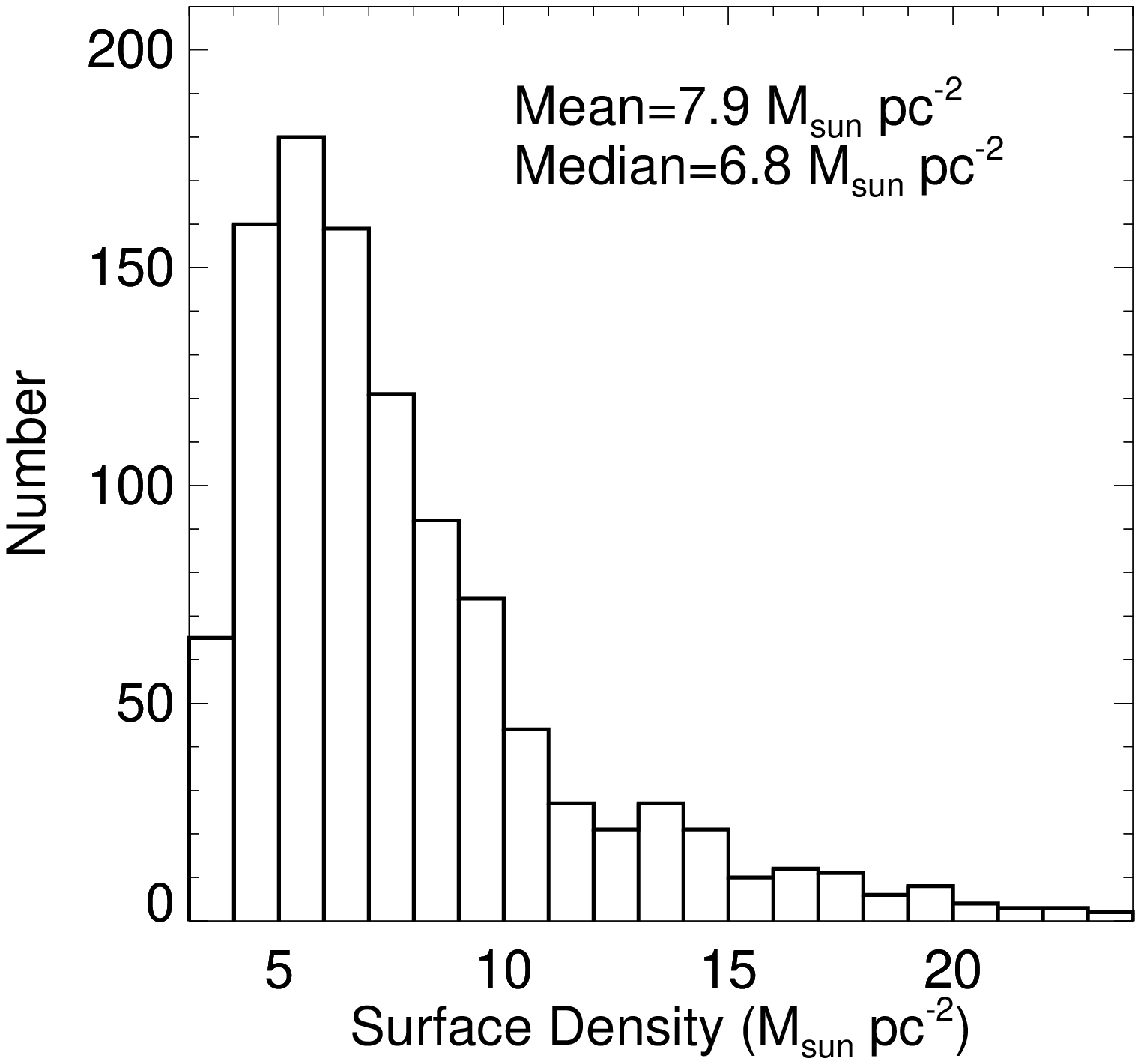}{0.35\textwidth}{(e)}
          \fig{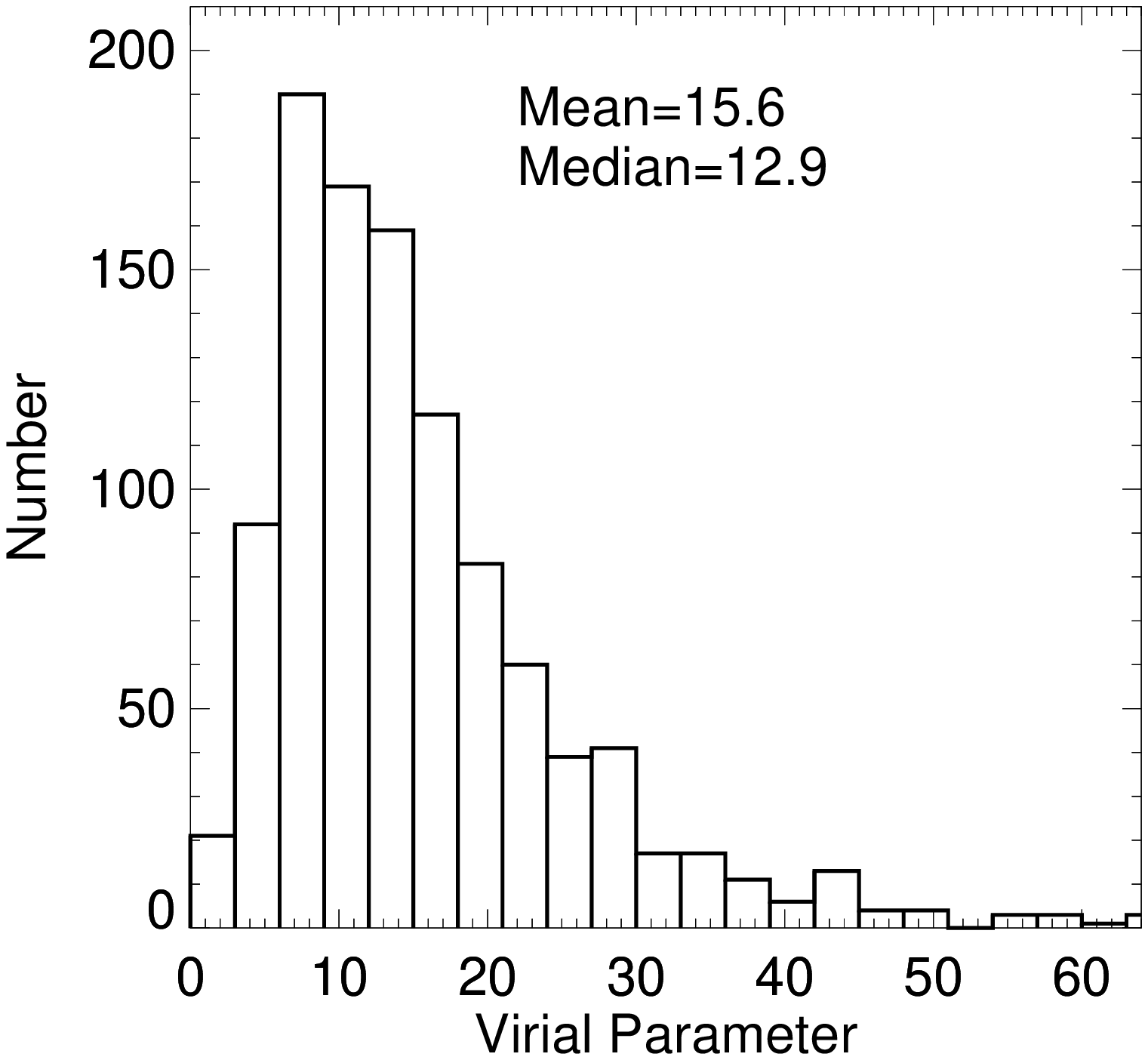}{0.35\textwidth}{(f)}
          }
\vspace{-2ex}
\caption{
(a)--(f) Histogram of the peak temperature, the velocity dispersion,
the radius, the mass, the surface density, and the virial parameter
of the \twCO\ cloud far from the Galactic plane. 
\label{sta_para}}
\end{figure}
\clearpage

\begin{figure}
\gridline{\fig{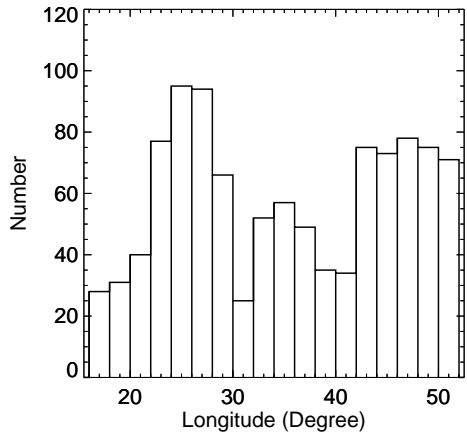}{0.35\textwidth}{(a)}
          \fig{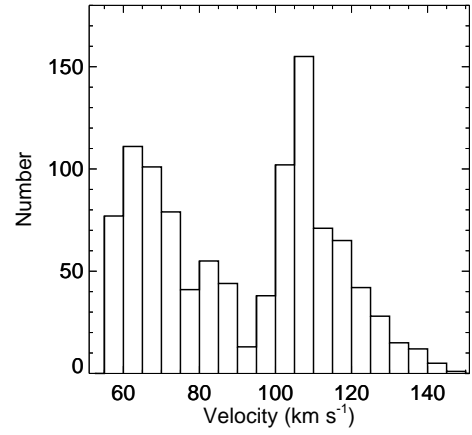}{0.35\textwidth}{(b)}
          }
\vspace{-5ex}
\gridline{\fig{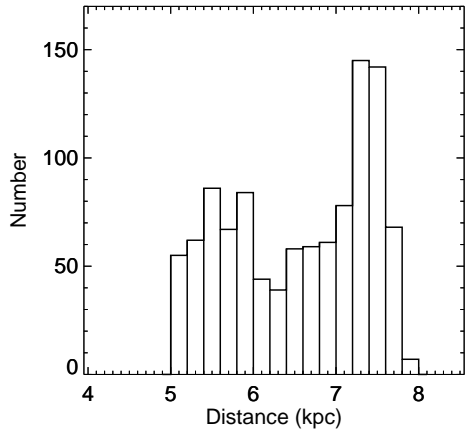}{0.35\textwidth}{(c)}
          \fig{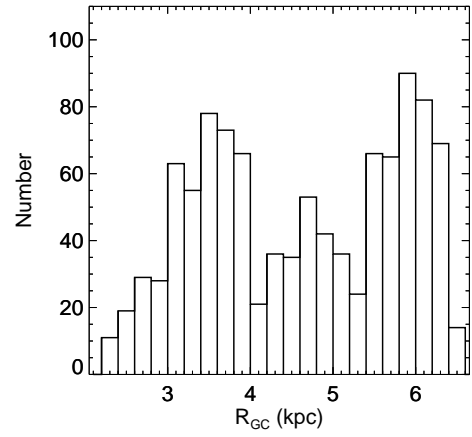}{0.35\textwidth}{(d)}
          }
\vspace{-2ex}
\caption{
(a)--(d) Longitude, velocity, distance, and Galactocentric distance
distributions of the MCs far from the Galactic plane.
\label{sta_z}}
\end{figure}
\clearpage

\begin{figure}
\includegraphics[trim=0mm 0mm 0mm 0mm,scale=0.7,angle=0]{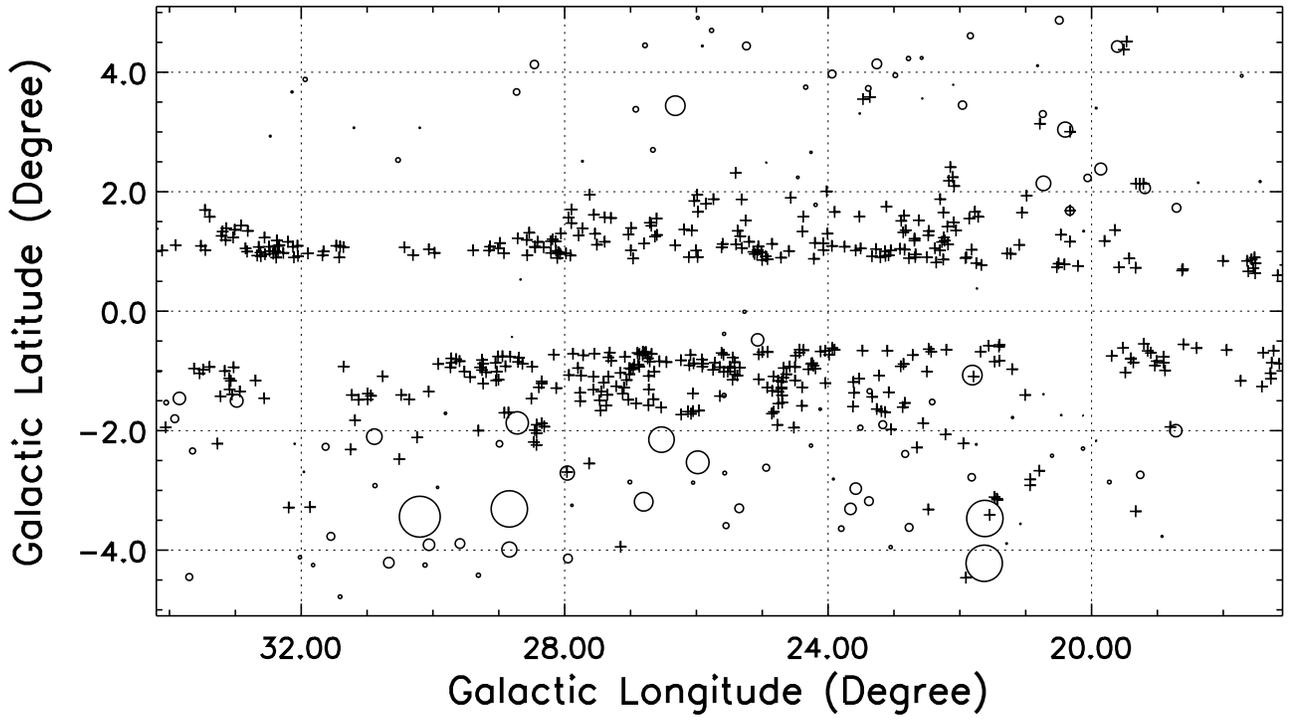}
\caption{
Comparison between the high-$z$ CO clouds (crosses with the uniform size; this paper) 
and the disk-halo \HI\ clouds \citep[circles;][]{Ford10}. Note that the 
circle's size is not the true angular size of the \HI\ clouds, but it is 
proportional to the mass of the \HI\ clouds.
\label{CO_HI}}
\end{figure}
\clearpage

\begin{figure}
\includegraphics[trim=10mm 0mm 0mm 0mm,scale=0.6,angle=0]{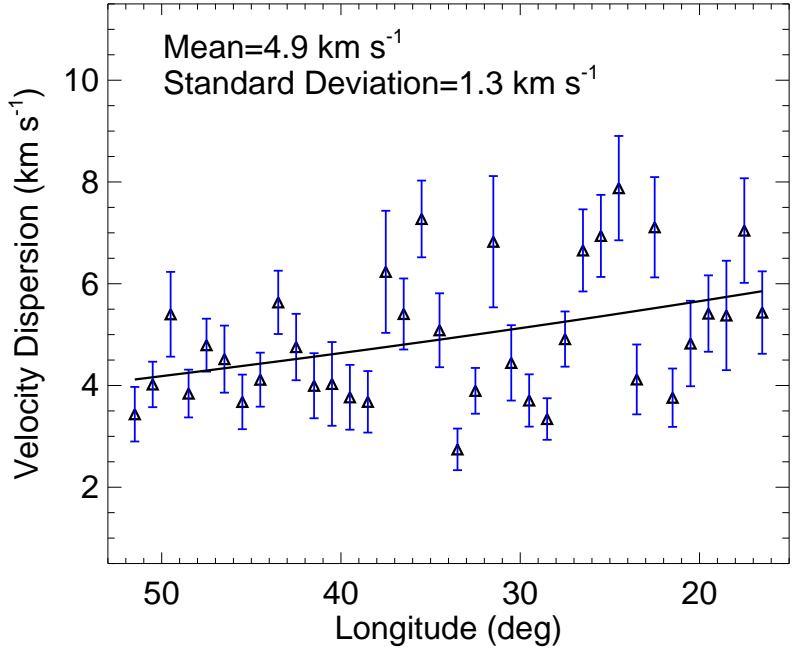}
\caption{
Cloud-cloud velocity dispersion ($\sigma_{\rm cc}$) for 
all MCs (i.e., the thin + thick CO disk)
near the tangent points toward $l=16^{\circ}-52^{\circ}$.
The line indicates the least-squares fit from a third-order polynomial
(nearly linear form with a slope of $\sim-0.4\ \km\ps$~kpc$^{-1}$) 
by using the current data of $R_{\rm GC}\sim$~2.2--6.4~kpc. The error 
for each bin of $1^{\circ}$ is assumed as $\sigma_{\rm cc}$/$\sqrt{N_{\rm cloud}}$.
The mean value of $\sigma_{\rm cc}$ and its standard deviation 
are also labeled on the panel.
\label{sigv_12}}
\end{figure}
\clearpage

\begin{figure}
\includegraphics[trim=10mm 0mm 0mm 0mm,scale=0.6,angle=0]{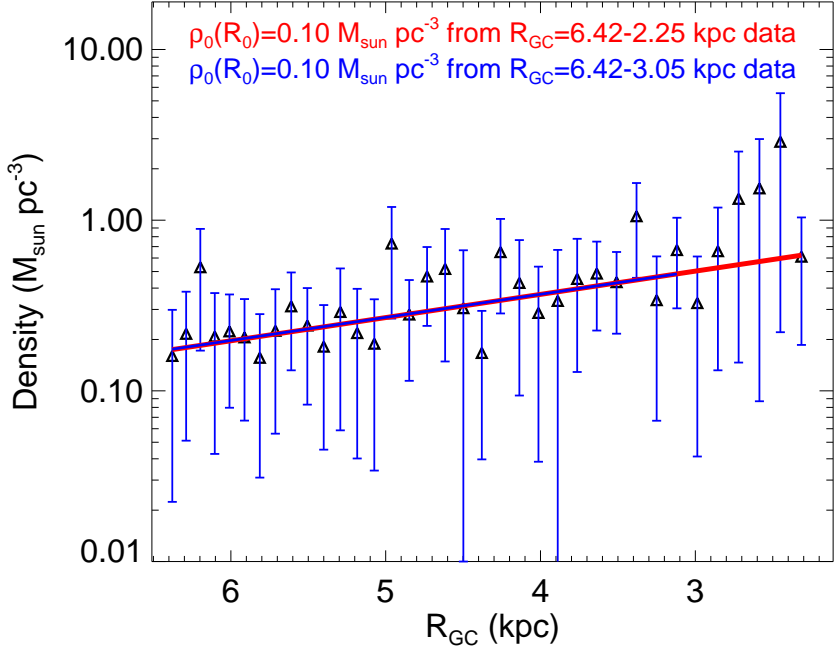}
\caption{
Midplane mass density for different Galactocentric distances ($R_{\rm GC}$) derived
from the CO scale height (Figure~\ref{z0width}b) and the velocity dispersion
(Figure~\ref{sigv_12}). The line shows the exponential-disk model of 
$\rho_0(R)=\rho_{\rm GC}e^{-R/R_{\rm sl}}$, where the fitted mass density at the 
Galactic center $\rho_{\rm GC}=1.28$~$\Msun$~pc$^{-3}$, and the scale length,
$R_{\rm sl}=3.20$~kpc. The midplane mass density at the Sun 
($R_{\rm GC}=R_0$=8.15~kpc) is estimated to be $\sim0.10\ \Msun$~pc$^{-3}$.
\label{density}}
\end{figure}
\clearpage

\clearpage
\begin{deluxetable}{ccccccccccc}
\tablecaption{Parameters of 1055 Molecular Clouds Far from 
the Galactic Plane based on the MWISP \twCO ($J$=1--0) Emission}
\tablehead{
\colhead{\begin{tabular}{c}
ID  \\
    \\
(1)  \\
\end{tabular}} &
\colhead{\begin{tabular}{c}
$l$               \\
($^{\circ}$)      \\
(2)  \\
\end{tabular}} &
\colhead{\begin{tabular}{c}
$b$               \\
($^{\circ}$)      \\
(3)  \\
\end{tabular}} &
\colhead{\begin{tabular}{c}
$V_{\rm LSR}$      \\
($\km\ps$)         \\
(4)  \\
\end{tabular}} &
\colhead{\begin{tabular}{c}
$\sigma_v$     \\
($\km\ps$)                  \\
(5)  \\
\end{tabular}} &
\colhead{\begin{tabular}{c}
$T_{\rm peak}$   \\
(K)              \\
(6)  \\
\end{tabular}} &
\colhead{\begin{tabular}{c}
Area             \\
(arcmin$^2$)     \\
(7)  \\
\end{tabular}} &
\colhead{\begin{tabular}{c}
Distance$^{\mathrm {a}}$            \\
(kpc)     \\
(8)  \\
\end{tabular}} &
\colhead{\begin{tabular}{c}
$z$ scale    \\
(pc)     \\
(9)  \\
\end{tabular}} &
\colhead{\begin{tabular}{c}
Mass           \\
($\Msun$)     \\
(10)  \\
\end{tabular}} &
\colhead{\begin{tabular}{c}
$\alpha^{\mathrm {b}}$     \\
     \\
(11)  \\
\end{tabular}}
}
\startdata
  1   &   16.103   &   0.896   &  143.16   &  2.07   &  1.84   &  10.75   &   7.8  &  122.5   &  690   &  31.5   \\
  2   &   16.278   &   0.976   &  137.99   &  1.00   &  1.29   &   2.50   &   7.8  &  133.2   &  100   &  24.2   \\
  3   &   16.436   &  -0.770   &  128.18   &  0.85   &  1.97   &   5.00   &   7.8  & -105.0   &  160   &  15.4   \\
  4   &   16.498   &   0.809   &  130.07   &  1.31   &  2.60   &  14.25   &   7.8  &  110.4   &  620   &  15.9   \\
  5   &   16.526   &   0.929   &  139.87   &  3.27   &  2.54   &  15.75   &   7.8  &  126.7   & 1160  &  56.4   \\
  6   &   16.768   &  -0.600   &  128.00   &  1.37   &  3.30   &  31.50   &   7.8  &  -81.8   & 1700   &   9.5   \\
  7   &   16.801   &  -1.173   &  128.58   &  0.77   &  1.70   &   6.50   &   7.8  & -159.8   &  220  &  10.5   \\
\hline
\enddata
\tablecomments{
$^{\mathrm {a}}$ The error of the estimated distance is about 2\%--20\% from 
$l=16^{\circ}-52^{\circ}$ assuming that the MCs are located near tangent points 
with a velocity uncertainty of $\sim5\km\ps$
along the line of sight \citep[e.g., refer to the A5 model in][]{Reid19}.
$^{\mathrm {b}}$ The MC's virial parameter estimated from the definition of
$\alpha=\frac{5\sigma_{v}^2R}{GM_{\rm Xfactor}}$ (see Section 3.3.1).
}
\end{deluxetable}

\begin{deluxetable}{cccccccccc}
\tablecaption{Statistical Properties of the high-$z$ Molecular Clouds$^{\mathrm {a}}$}
\tablehead{
\multicolumn{6}{c}{Median} & & \multicolumn{3}{c}{Mean} \\
\cline{1-6} \cline{8-10} \\
\colhead{\begin{tabular}{c}
$T_{\rm peak}$  \\
(K)                    \\
\end{tabular}} &
\colhead{\begin{tabular}{c}
Radius               \\
(pc)                        \\
\end{tabular}} &
\colhead{\begin{tabular}{c}
$\sigma_v$        \\
($\km\ps$)             \\
\end{tabular}} &
\colhead{\begin{tabular}{c}
Mass      	 \\
($\Msun$)        	 \\
\end{tabular}} &
\colhead{\begin{tabular}{c}
Surface Density         \\
($\Msun\ {\rm pc}^{-2}$)                \\
\end{tabular}} &
\colhead{\begin{tabular}{c}
$\alpha^{\mathrm {b}}$           \\
                   \\
\end{tabular}} &
\colhead{\begin{tabular}{c}
		   \\
                   \\
\end{tabular}} &
\colhead{\begin{tabular}{c}
$\sigma_{\rm cc}$          \\
($\km\ps$)                      \\
\end{tabular}} &
\colhead{\begin{tabular}{c}
Thickness      \\
(pc)              \\
\end{tabular}} &
\colhead{\begin{tabular}{c}
$\gamma^{\mathrm {c}}$      \\
            \\
\end{tabular}} 
}
\startdata
 2.1  &   2.5   &  0.8  &  136.3   &  6.8  & 12.9  &  & $\sim$4.4--5.6  & $\sim$280  & -1.74 \\
\hline
\enddata
\tablecomments{
$^{\mathrm {a}}$ The properties of the molecular gas indicate that the high-$z$ MCs 
belong to a new disk population that was not discovered by previous CO observations
due to the low sensitivity, low resolution, and/or limited latitude coverage.
$^{\mathrm {b}}$ $\alpha$ is the virial parameter of the MCs (see the text).
$^{\mathrm {c}}$ $\gamma$ is the spectral index of the mass distribution of 
the high-$z$ MCs with the form $\frac{dN}{dM}\propto M^{\gamma}$
in the MC's mass range of $\sim120-8000\ \Msun$.
}
\end{deluxetable}

\begin{deluxetable}{cccc}
\tablecaption{Thickness of the Inner Galaxy Traced by Different Tracers}
\tablehead{
\colhead{Tracer} & \colhead{FWHM$^{\mathrm {a}}$}   & 
\colhead{References$^{\mathrm {b}}$} & \colhead{Comments}}
\startdata
CO     &  $\sim$85~pc    &  1,2  &  The well-known thin molecular gas disk \\
       &  $\sim$280~pc   &  1    &  The thick disk is composed of many discrete MCs with small size and low mass\\
\hline
\CII\  &  $\sim$120~pc   &  3    &  \CII\ with bright CO emission as the dense H$_2$ gas (the overestimated thin CO disk)\\
       &  $\sim$190~pc   &  3    &  Bright diffuse \CII\ emission as the CO-faint diffuse H$_2$ gas \\
       &  $\sim$320~pc   &  3    &  Faint diffuse \CII\ emission as the diffuse \HI\ and WIM  gas \\
\hline
\HI\   &  $\sim$230~pc   &  4,5  &  The narrow Gaussian component for the cold neutral medium \\
       &  $\sim$540~pc   &  4,5  &  The broad Gaussian component for the warm neutral medium \\
       &  $\sim$1300-1600~pc  &  5,6    &  The exponential component for the warm neutral medium (the disk-halo gas)\\
\hline
\enddata
\tablecomments{
$^{\mathrm {a}}$ The estimated thickness has been scaled to $R_0=$8.15~kpc
\citep[e.g.,][]{Reid19}.
$^{\mathrm {b}}$ 1. This paper, 2. \cite{Malhotra94}, 3. \cite{Velusamy14}, 
4. \cite{Dickey90}, 5. \cite{Lockman91}, 6. \cite{Ford10}.}
\end{deluxetable}

\end{document}